\documentclass{aa}
\usepackage{txfonts}
\usepackage{natbib}
\RequirePackage{graphicx}%

\begin{document}

\title{X-ray Detection of the Proto Supermassive Binary Black Hole at the Centre of Abell 400}

\author{D. S. Hudson\inst{1}\thanks{Email Address:dhudson@astro.uni-bonn.de}
\and T. H. Reiprich \inst{1}
\and T. E. Clarke\inst{2,3} 
\and C. L. Sarazin\inst{4}}

\institute{Argelander-Institut f\"ur Astronomie der Universit\"at Bonn, Auf dem H\"ugel 71, D-53121 Bonn, Germany
\thanks{Founded by the merging of the Institut f\"ur Astrophysik und Extraterrestrische Forschung, the Sternwarte, and the Radioastronomisches Institut der Universit\"at Bonn.}
\and Naval Research Laboratory, 4555 Overlook Ave. SW, Code 7213, Washington, DC, 20375, USA
\and Interferometrics Inc., 13454 Sunrise Valley Drive, Suite 240, Herndon, VA, 20171, USA
\and Department of Astronomy, University of Virginia, P. O. Box 3818, Charlottesville, VA 22903-0818, USA}
\date{03-02-2006 / 09-03-2006}

\abstract
{We report the first X-ray detection of a proto-supermassive binary black hole  at the centre of Abell 400.  Using the {{\it Chandra}} {{\it Advanced CCD Imaging Spectrometer}}, we are able to clearly resolve the two active galactic nuclei in 3C 75, the well known double radio source at the centre of Abell 400.}
{Through analysis of the new {{\it Chandra}} observation of Abell 400 along with 4.5 GHz and 330 MHz {{\it Very Large Array}} radio data, we will show new evidence that the Active Galactic Nuclei in 3C 75 are a bound system.}
{Using the high quality X-ray data, we map the temperature, pressure, density, and entropy of the inner regions as well as the cluster profile properties out to $\sim18\arcmin$.  We compare features in the X-ray and radio images to determine the interaction between the intra-cluster medium and extended radio emission.}
{The {{\it Chandra}} image shows an elongation of the cluster gas along the northeast-southwest axis; aligned with the initial bending of 3C 75's jets.  Additionally, the temperature profile shows no cooling core, consistent with a merging system.  There is an apparent shock to  the south of the core consistent with a Mach number of ${\cal M}\sim1.4$ or speed of $v\sim$1200 km s$^{-1}$. Both Active Galactic Nuclei, at least in projection, are located in the low entropy, high density core just north of the shock region.  We find that the projected path of the jets does not follow the intra-cluster medium surface brightness gradient as expected if their path were due to buoyancy.  We also find that both central AGN are extended and include a thermal component.}
{Based on this analysis, we conclude that the Active Galactic Nuclei in 3C 75 are a bound system from a previous merger.  They are contained in a low entropy core moving through the intra-cluster medium at 1200 km s$^{-1}$.  The bending of the jets is due to the local intra-cluster medium wind.}


\maketitle

\section{Introduction}
The production and coalescence of super-massive binary blackholes ({\it SMBBHs}) seems to be a natural consequence of galaxy mergers.  Specifically, when the large central galaxies of clusters merge, such binaries should be formed.  The decay of a {\it SMBBH} goes through three stages: (1) energy loss via dynamical friction, (2) energy loss through ejection of stars, and (3) energy loss through emission of gravity waves \citep[e.g.][and references there in]{milosavljevic03}.  The final stage has received a lot of attention lately because it may be the largest source of gravitational waves detectable by Laser Interferometer Space Antenna ({\it LISA}) \citep[e.g.][]{merritt04}.  The understanding of how {\it SMBBHs} form and coalesce is important for understanding Active Galactic Nuclei (AGN) dynamics as well as galaxy formation \citep[e.g.][]{komossa03b}.  Although there has been no direct observation of a {\it SMBBH} with a separation under 1 kpc \citep{merritt04}, it is possible for {\it Chandra} to make such a detection in a nearby galaxy.  Recently \citet{komossa2003} discovered a {\it SMBBH} with a projected separation of $\sim$1.4 kpc, at the centre of NGC 6240.  Although this source had already been identified in optical observations, the {\it Chandra} observation was needed to confirm that both sources were AGN. 

\object{3C 75} is the well-known Wide Angle Tail ({\it WAT}), double radio source at the centre of nearby galaxy cluster \object{Abell 400} (A400).  Early radio observations by \citet{owen85}  first revealed the dual source nature of 3C 75, and provided details about the radio jets.  These authors concluded that the evidence for interaction between the jets suggested that the two sources were physically close, and the proximity was not simply a projection effect.  Optical observations by \citet{hoessel85} implied that the dumbbell galaxy hosting the two sources was in fact two galaxies which were not interacting.  However, in a later optical analysis, \citet{lauer88} argued the contrary.  Additionally, \citet{yokosawa85} modelled the tails of the extended radio emission, demonstrating that their morphology could be explained if the radio lobes were interacting while the cores orbited each other.  By comparing the 4.9 GHz radio image by \citet{owen85} to simulations of intertwining jets, \citet{yokosawa85} derived several parameters which suggest that 3C 75 is a bound system that is rapidly losing energy.  They also estimate the actual separation of the blackholes to be $\sim$8 kpc (compared to the 7 kpc separation seen in projection), although they admit that their simplified model probably underestimates the actual separation.  They also conclude that the southern source (A400-218 in their paper) is more massive by a factor of $\sim$2, and its galaxy has not yet been tidally disrupted by the northern source (A400-217 in their paper).  Their simulations predict that a relative speed of 1120 km s$^{-1}$ between 3C 75 and the intracluster medium (ICM) is needed to produce the bending seen on the large scale.

Based on an early {\it Einstein} X-ray observation presented by \citet{forman82}, \citet{yokosawa85} suggested that A400 was relaxed due to its apparent spherical symmetry.  This led them to believe that ICM wind was not strong enough to account for the bending of the jets seen on the large scale.  In fact, both \citet{yokosawa85} and \citet{owen90} suggest that A400 is relaxed, so that any local ICM wind would be insufficient to bend the jets.  Based on this supposition, \citet{owen90} used long slit analysis to determine if the bending of the jets in 3C 75 was due to cool dense clouds in the ICM.  Although they could not rule out the possibility of cool dense clouds, they concluded that it was very unlikely since the parameters needed required rather hot clouds ($T$ $>$ 10$^5$ K) and supersonic jet velocities ($v$ $\sim$ 10$^4$ km s$^{-1}$).

A detailed analysis of the galaxies by \citet{beers92} revealed at least two separate subclusters which make up A400.  According to their analysis, the two subclusters lie (in projection) on top of each other.  They suggest either a near line of sight merger or that the subclusters are crossing each other.  
In addition, these authors present the same {\it Einstein} X-ray data as \citet{forman82} and conclude that the X-ray emitting gas is elongated along the same direction as the initial bending of the jets.  They claim that based on their merger scenario and the elongation of the X-ray emitting ICM, the relative motion of the two subclusters could explain the initial bending of the jets due to the relative motion of the AGN through the ICM.  In this case, since the bending is in the same direction, a bound system is suggested since both the AGN would have to have the same relative motion in order for their jets to be bent in the same direction.

\citet{beers92} in their analysis for a bound AGN system give their best fit model as a merging system with a projection angle of $\sim$18$^{\circ}$.  In this case, the velocity difference between the two sub-clusters is $\sim$2000 km s$^{-1}$ (based on their measured radial velocity difference between the subclusters of $\sim$700 km s$^{-1}$).  In this scenario the velocity of the merger is enough to explain the initial bending of the jets.  They also point out that the two components of the dumbbell galaxy have velocities (6800 km s$^{-1}$ (Northern Source) and 7236 km s$^{-1}$ (Southern Source) \citep{Davoust95}) that put them close to the middle of the velocity distribution of each subcluster (6709 km s$^{-1}$ and 7386 km s$^{-1}$). Based on this, they suggest that the dumbbell galaxy may be the remnant of the two dominant galaxies of the subclusters.

In this paper we explore radio and new $Chandra$ X-ray data to analyse the structure of the ICM and to study the nature of 3C 75.  Specifically, we study the extended radio emission and compare it to parameters of the ICM and try to determine whether 3C 75 is a bound system.

\section{Observations and Methods}
A400 is a nearby cluster (z = 0.0244 \citep{struble99}, 1$\arcmin$ = 29.1  $h_{71}^{-1}$ kpc), located at $\alpha$(J2000) = 02$^{h}$57$^{m}$39.7$^{s}$, $\delta$(J2000)= +06$^{\circ}$01$\arcmin$01$\arcsec$ \citep{reiprich02}.  3C 75 is a {\it WAT} with a double core and intertwining jets, located at the centre of A400.  The northeast AGN is at a redshift of $z$ = 0.022152, while the southwest AGN is at a redshift of $z$ = 0.023823 \citep{Davoust95}.  Adopting the nomenclature of \citet{dressler80}, we label the northern AGN {\it A400-42} and southern AGN {\it A400-43}.  Our calculations are done assuming a flat $\Lambda$CDM universe with $\Omega_{M}$=0.3, $\Omega_{\rm vac}$=0.7, and H$_{0}$=71 $h_{71}$ km s$^{-1}$ Mpc$^{-1}$.  All errors are quoted at the 90\% level unless otherwise noted.

\subsection{X-ray Data Reduction}
\label{datred} 
A400 was observed with $Chandra$ on 2003-Sep-19 for $\sim$22 ks, as part of the Highest X-ray Flux Galaxy Cluster Sample ({\it HIFLUGCS}) \citep{reiprich02} follow-up program.  We reduced the data using the routines from CIAO 3.2.2 Science Threads\footnote{{\tt http://cxc.harvard.edu/ciao/threads/index.html}} and calibration information from CALDB 3.1.0.  More specifically as advised, we followed the procedures in M. Markevitch's cookbook.\footnote{{\tt http://cxc.harvard.edu/cal/Acis/Cal\_prods/bkgrnd/acisbg/COOKBOOK}}  We included all 5 Chips (I0, I1, I2, I3, and S2) in our analysis, but reduced the data in the I-Chips separately from the S2 Chip.

We started from the level-1 events in order to take advantage of the latest calibration data and software.  We created an observation-specific bad pixel file using {\it acis\_run\_hotpix}.  We used the very faint mode grades to remove extra background.  To be consistent with the background files we de-streaked the Chips (all Chips are front illuminated), and did not remove pixels and columns adjacent to bad pixels and columns (see M. Markevitch's cookbook at address listed above). 

A400 fills the entire field of view of {\it Chandra}, including the S2 Chip.  Therefore, we used the latest 1.5 Ms background files produced by M. Markevitch\footnote{{\tt http://cxc.harvard.edu/cal/Acis/Cal\_prods/bkgrnd/acisbg/data/}} as blank-sky background.  These background files have their own set of bad pixels, so we combined our bad pixels with those from the background and used A. Vikhlinin's {\it badpixfilter} script (found on M. Markevitch's web-site listed above) to filter the events files and background files on the combined set.  We filtered the background file with all status bits = 0, in order to make them equivalent to our very faint observation.  Basically, all the bits have been set to zero except for status bit 9, which are events which have been flagged by {\it check\_vf\_pha}.  Finally, we reprojected the background into source sky coordinates.

We used {\it wavdetect} on scales of 1.0, 2.0, 4.0, 8.0, and 16.0 pixels, to detect sources within the field of view.  We edited these sources where the algorithm underestimated the width of several point-like sources.  For analysis of the ICM we excluded the two central AGN (3C 75), but obviously did not remove them when analysing 3C 75.  Once the point sources were removed, we filtered the light curve for flares using A. Vikhlinin's algorithm {\it lc\_clean}.  We filtered over all four I-Chip GTIs so that we could combine all four I-Chips and bin the light curve in smaller intervals (259.28 s bins).  We used the default settings for the I-Chips (0.3 - 12 keV, 3-sigma clip to calculate the mean, and 20\% cut above and below the mean).  After filtering the lightcurve for the I-Chips, we were left with $\sim$21.5 ks of data.  For the S2 Chip, we used a larger bin size (1037.12 s) since we were only filtering a single chip, but kept all other parameters the same.  In the case of the S2 Chip, we were left with $\sim$19.7 ks after filtering the lightcurve.

Source emission in the 9.5-12 keV range is dominated by background, so we normalised the blank-sky background to the source rate in this band.  The count rate in this range for the I-Chips suggests that the background in our observation is $\sim$95.3\% of the blank-sky background.  For the S2 Chip, the normalisation factor is $\sim$98.6\%.

Using the script {\it make\_readout\_bg} we created a so-called Out-of-Time (OOT) (also known as Readout Artifact) events file from our level 1 events file.  We processed the OOT events identically to the observation events.  Using the method outlined in M. Markevitch's script {\it make\_readout\_bg}, we normalized the OOTs to account for the $\sim$0.013 ratio between the read-out time and the total observation time.

\subsection{Image Creation}
\label{imgcre}
\begin{figure}
\includegraphics[width=85.0mm]{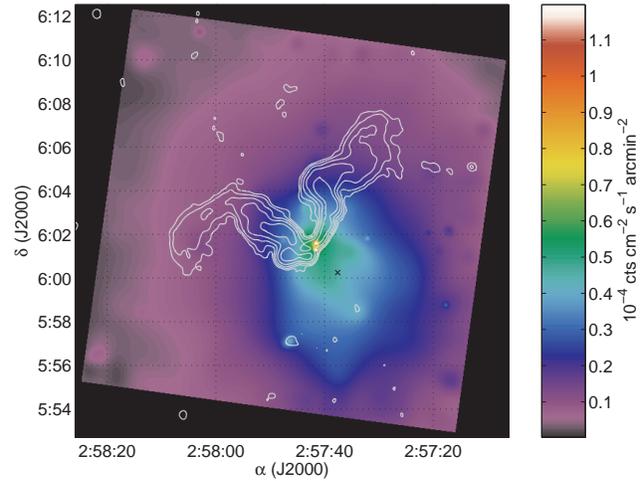}
\caption[A400: {\it ACIS} X-Ray Surface Brightness Image with 325MHz Radio Contour Overlay]
{Adaptively smoothed X-ray surface brightness map of A400 in the 0.7-7.0keV range with an overlay of VLA 329MHz contours. The colour scale has been cut at 1.2 $\times$ 10$^{-4}$ cnts cm$^{-2}$ s$^{-1}$ arcmin$^{-2}$ to prevent 3C 75 from dominating the scale of the image.  The diffuse X-ray, emission weighted centre of the cluster is labelled with an {\it x} and is $\sim$1$\arcmin$.68 southwest of the emission peak.  The radio contours are logarithmically scaled from 3$\sigma$ to the peak signal. \label{fig-A400-SBMap}}
\end{figure}

The effective area of {\it Chandra} is a function of both energy and position.  Exposure maps can be created to take into account the spatial effects but can only be created for a single energy.  Therefore, we created our {\it Chandra} surface brightness map (see Fig.-\ref{fig-A400-SBMap}) by combining surface brightness maps from 20 unique energy bands, each with an equal number of counts.  

\begin{figure}
\includegraphics[width=85.0mm]{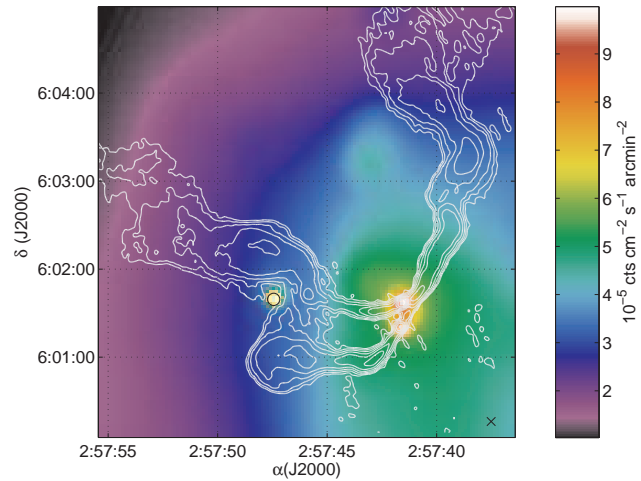}
\caption[3C 75: {\it ACIS} X-Ray Surface Brightness Image with VLA Radio Contour Overlay]
{A zoom-in of the central region of Fig.-\ref{fig-A400-SBMap} with an overlay of the VLA 4.5 GHz contours.  The diffuse X-ray, emission weighted centre is labelled with an {\it x}. The X-ray emission from {\it A400-41} has not been removed from this image and its centre is marked with an {\it o}.  Note that in projection, {\it A400-42}'s eastern radio lobe appears to be deflected and travel around {\it A400-41} (see Sect.-\ref{ejet}). The AGN X-ray emission from 3C 75 has been removed from this image.  The radio contours logarithically spaced from 3$\sigma$ to the peak signal. \label{fig-A400-C1}}
\end{figure}

We smoothed the raw (0.7 - 7.0 keV) image to create a smoothing kernel for each pixel using the {\it CIAO} tool {\it csmooth}.  We used this adaptive kernel to smooth the source image, background image, OOT image, and exposure map of each band.  Then, we subtracted the background and OOTs from the smoothed data, divided the result by the smoothed exposure map, and finally combined the 20 bands to form a single image (see Fig.-\ref{fig-A400-SBMap} and Fig.-\ref{fig-A400-C1}).

\subsection{Annular Regions}
\label{annreg}
In order to model the annular properties of A400, we first found the emission weighted centre (EWC) of the {\it Chandra} image.  We used a 6$\arcmin$ circle to iteratively (starting from the {\it HIFLUGCS} centre) determine the EWC on a background subtracted, exposure corrected, unsmoothed image.  We found the final centre to be, $\alpha$(J2000) = 02$^{h}$57$^{m}$37.53$^{s}$, $\delta$(J2000)= +06$^{\circ}$00$\arcmin$16$\arcsec$, and took this to be the centre of each annulus.  We then chose each annulus so that it contained at least 5000 source\footnote{Source means background and OOTs have been subtracted and point sources removed.} counts in the energy range we used to fit our spectra (0.6-10 keV).  The penultimate region, five, was taken to be a partial annulus that extended almost to the edge of the chip, in order to have as large a radius as possible and still remain on the I-Chips.  This region has $\sim$5300 source counts in the 0.7-10 keV range and a signal to noise ratio of $\sim$49.9 (compared to $\sim$56.1 for Region-4).  The outermost region, six, was concentric, but since it is extracted from the S2 Chip, it is not adjacent to the other annuli. This region was treated separately from all the other regions.  Region six is a partial annulus that contains as much emission from the S2 Chip as possible, without any section of region extending beyond the edge of the chip.  It contains $\sim$500 source counts with a signal to noise ratio of $\sim$9.6.

We extracted a spectrum for each annulus and the total cluster emission on the I-Chips. We corrected each spectrum using a corresponding blank-sky background.  Additionally, for the five (5) inner annuli, we corrected for $OOT$ events.  For the outer most annulus (the region on the S2 Chip) and the total cluster spectrum, we did not correct for $OOT$ background.  These regions contain the majority of the emission on the corresponding Chip(s) and so the misplacement of events during read-out does not affect the spectrum.  We binned each source spectrum to at least 25 counts per bin with errors taken to be Gaussian so they could be added correctly in quadrature.

We fit each spectrum over the 0.6-10.0 keV range to an absorbed $APEC$ \citep{smith01a,smith01b,smith00} model with an additional component to account for the incorrect modelling of $Chandra$ effective area above the Ir-M edge \citep{vikhlinin05}.  This component is well modelled by an $edge$ parameter:
\begin{equation}
A = exp(-\tau(E/E_{\rm thresh})^{-3}) \; {\rm for}\:E\:>\:E_{\rm thresh},
\end{equation}
where $E_{\rm thresh}$ = 2.07 keV and $\tau$=-0.15 (a so called $positive$ $absorption$) \citep{vikhlinin05}. For the photoelectric absorption we used the Wisconsin cross-sections \citep{morrison}.  As discussed in detail below (Sect.-\ref{phoabs}), we found some a discrepancy in some regions between the column densities determined by radio measurements ($N_{H}$ = 8.51 $\times$ 10$^{20}$ cm$^{-2}$ \citep{kalberla05}) and the best fit to the X-ray data.  Therefore we fit models with both free and frozen absorption models.  For the overall cluster spectrum, in addition, to a single thermal model, we also fit the overall cluster emission to a double thermal model.  

\subsection{Temperature Map}
\label{tmap} 
\begin{figure}
\includegraphics[width=85.0mm]{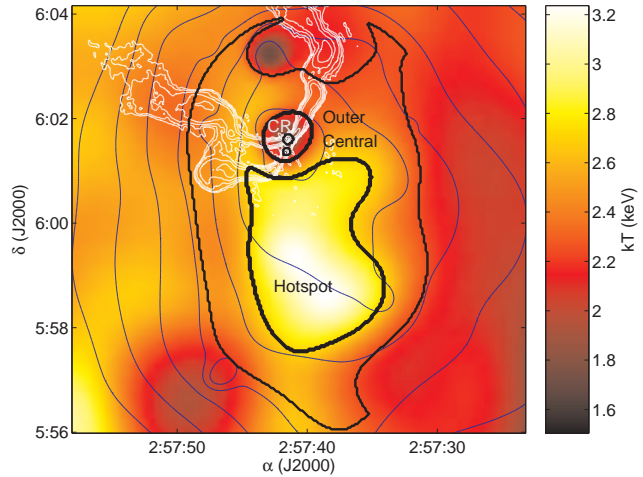}
\caption[A400: {\it ACIS} Hardness Implied Temperature Map]
{The hardness implied temperature map of the central region of A400 with X-ray surface brightness contours.   This map was made by mapping hardness ratios to temperatures of a {\it MEKAL} model with the best fit cluster wide parameters for photoelectric absorption ($N_{H}$ = 9.97 $\times 10^{20}$ cm$^{-2}$) and metalicity (Z = 0.52 Solar).  The special fitted regions are labelled: CR-Central Region, Outer Central Region, and Hotspot Region.  Note that the Central Region excludes the two AGN.  The 4.5 GHz radio contours are identical to those in Fig.-\ref{fig-A400-C1} \label{fig-A400-kTMap}}
\end{figure}

Hardness ratios can give an estimation of the temperature of the gas in a specific region.  We created two hardness maps by creating images identical to the method described in Sect.-\ref{imgcre}, but using a larger smoothing kernel (a significance level of 5$\sigma$ rather than 3$\sigma$) for better statistics.  These hardness maps were made by dividing our 20 images  into 2 sets.  We chose two hardness ratios with cuts at 1.27 keV and 1.34 keV, so that the central regions' hardness ratios were close to one.  For both hardness cuts we created a look up table which assigned a temperature for a given hardness ratio.  We used  an absorbed {\it MEKAL} \citep{mewe85,mewe86,kaastra92,liedahl95} model with our best fit overall cluster metalicity and photoelectric absorption to generate the look up table.  The resulting hardness implied temperature map (see Fig.-\ref{fig-A400-kTMap}) was created by averaging the resulting temperatures from each hardness map, and throwing out any points in which they differed by more than 10\%, or the surface brightness was less than 10$^{-5}$ counts cm$^{-2}$ s$^{-1}$ arcmin$^{-2}$.

Various techniques have been employed to map the complex structure of the ICM in clusters of galaxies.  With {\it Chandra} and {\it XMM-Newton}, it is now possible to make true two-dimensional plots of the projected temperature structure.  In order to extract interesting regions, we cross-correlated the hardness and surface brightness maps to find regions of similar temperature and density.  For our observation,  the central regions had enough counts to do spectral studies.  The three regions we extracted from the central region of A400 are shown in Fig.-\ref{fig-A400-kTMap}.  The {\it Hotspot} (HS) appears to have a higher temperature than the surrounding ICM, whereas the {\it Central Region} (CR) seems to be a cool spot.  The final region, {\it Outer Central} (OC), was extracted to provide a statistical comparison of the gas around the HS and CR.  We extracted spectra and modelled the spectra from these regions identically to the method described for the annuli in Sect.-\ref{annreg}.

\subsection{Deprojection}
\label{deproj} 
The properties (especially density) of central regions are affected by the projection of the gas in the outer regions on to them.  For the annular regions, co-fitting the spectra with the correct constant factors can be used to model deprojected spectra.  Currently a model, $PROJCT$, exists in $XSPEC$ to do this type of deprojection,  however it assumes that the Galactic column density is the same in all regions.  We discuss this problem as well as our solution in detail in Sect.-\ref{phoabs}.  

For oddly placed and shaped regions, such as the $HS$, the $CR$, and the $OC$ region, deprojection is more difficult.  In order to deproject these, we have to make three assumptions: (1) that the region of interest is located on the plane of the sky which passes through the centre of the cluster, (2) that the region can be approximated as a sphere or a torus, and (3) that the emission projected on to the region is well represented by concentric annuli.  The argument for the first assumption is that the feature that makes the region interesting, by definition, dominates the spectrum, therefore its most likely position is in the densest part of the cluster along the line of sight.  The second assumption is needed to estimate the extent of the region along the line of sight.  In the case of the HS and CR, both of which contain no major holes, we approximated them as spherical regions.  We calculated the radius such that the area of the circle with the given radius has the same area as the region.  That is, the radius of the approximated sphere is $r = \sqrt{A/\pi}$, where $A$ is the area of the region.  The centre of the sphere is taken to be the geometric centre of the region.  For the OC region, we approximated it as a torus.  In projection, the torus is seen as an annulus with an outer radius, $r_{\rm out}$ and an inner radius, $r_{\rm in}$.  The centre of this annulus is taken to be the geometric centre of the region.  The mean radius, $r = (r_{\rm out} + r_{\rm in})/2$, of the annulus is taken to be the average distance of the points in the region from the region's centre.  The width of the annulus ($r_{\rm out} - r_{\rm in}$)  is taken such that the area of the annulus has the same area as the region.  Finally, in order to deproject, we need to have information about the gas projected on to the region.  The annuli provide an emission weighted average of the properties of the gas at a given radius from the EWC.  Since we have no way to correlate the emission in the plane of the sky at a certain distance from the region to the emission along the line of sight at the same distance, we argue that a statistical average of the gas at a given radius is the best estimation.

As a practical matter, we point out that the annuli lie on different parts of the detector than the region of interest.  Since the response of the detector varies with position, the emission from the annuli can not be directly subtracted off from a region of interest, but must be co-fit with it.  In order to determine the contribution of a particular shell (associated with an annulus) to a given region, we numerically calculated the volume of each concentric shell penetrated by the line of sight cylinder, $V_{\rm int}$.  When co-fitting the data, we use the normalisation for the shell, so that the fraction of emission projected onto the region of interest is simply $V_{\rm int}$ divided by the volume of the corresponding shell.  Additionally, this calculation had to be done for the concentric annuli.  

To co-fit the data we needed the same number of absorbed {\it APEC} models as we had data sets.  Each individual absorbed {\it APEC} model was tied across data sets, but each had a different constant representing its contribution to that particular data set/region.  

\subsection{Derived Parameters}
\label{derpar} 
Using the parameters of the spectral fits to the regions, we derived approximate values for the pressure and entropy for the regions.  We define entropy\footnote{The classical definition of entropy is $S = 3/2 k(N_{e} + N_{i}) ln(T/\rho^{2/3})$ + const. for $\gamma=5/3$.} as K = $kT/{n_{e}}^{2/3}$ and pressure as P = $kTn_{e}$.  The temperature, $kT$ is taken directly from the temperature measurements, including uncertainty.  We assumed that the electron density, $n_{e}$, across a given annulus is constant and determined it from the normalisation of the {\it APEC} model:
\begin{equation}
N = \frac{10^{-14} }{4 \pi (D_{A}\;(1+z))^{2}}\int n_{e} n_{H} dV,
\label{normeqn}
\end{equation}
where $N$ is the normalisation, $D_{A}$ is the angular size distance, $z$ is the redshift, $n_{e}$ is the electron number density, $n_{H}$ is the hydrogen number density, and $dV$ is the volume element.

In order to determine the uncertainty in entropy and pressure we had to take into account the uncertainty in temperature and normalisation.  Since these errors are not Gaussian, it is not possible to add them in quadrature.  We constructed a 90\% level confidence contour by stepping through the allowed values of temperature and normalisation and interpolating points between.  We then calculated the pressure and entropy at 1000 points along the contour.  We took the uncertainty in pressure and entropy to be the greatest and least value allowed by the contour.

\section{Results and Analysis}
\subsection{The Two AGN}
Using this {\it Chandra} observation, we determined {\it A400-42}'s X-ray centre as $\alpha$(J2000) = 02$^{h}$57$^{m}$41.56$^{s}$, $\delta$(J2000)= +06$^{\circ}$01$\arcmin$36.6$\arcsec$ and {\it A400-43}'s X-ray centre as $\alpha$(J2000) = 02$^{h}$57$^{m}$41.63$^{s}$, $\delta$(J2000)= +06$^{\circ}$01$\arcmin$20.5$\arcsec$.  Using their average redshift, z = 0.023513, this gives them a projected separation of 7.44 $h_{71}^{-1}$ kpc.  This is the second closest projected distance of two potentially bound {\it SMBH} detected in X-rays.

\begin{figure}
\includegraphics[width=85.0mm,angle=0]{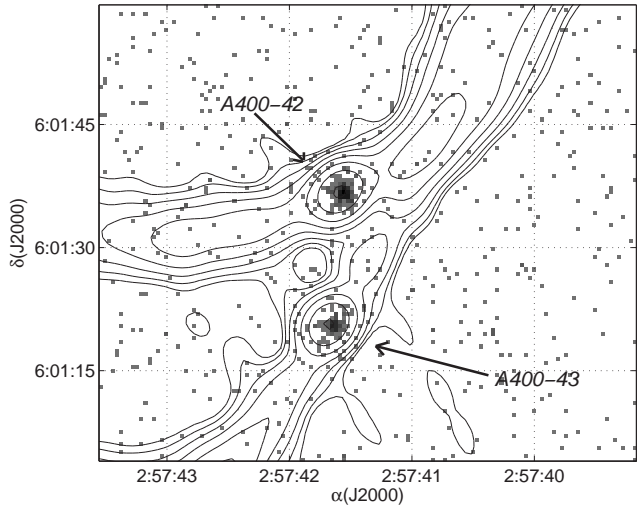}
\caption[3C 75: {\it ACIS} X-Ray Surface Brightness Image with VLA Radio Contour Overlay]
{A zoom-in of the raw {\it Chandra} image of the central region of A400, with an overlay of VLA 4.5 GHz contours. The double nucleus of 3C 75 is clearly separated.  The radio contours are logarithmically spaced from 3$\sigma$ to the peak signal.  \label{fig-A400-Zoom}}
\end{figure}

\begin{figure}
\includegraphics[width=85.0mm,angle=0]{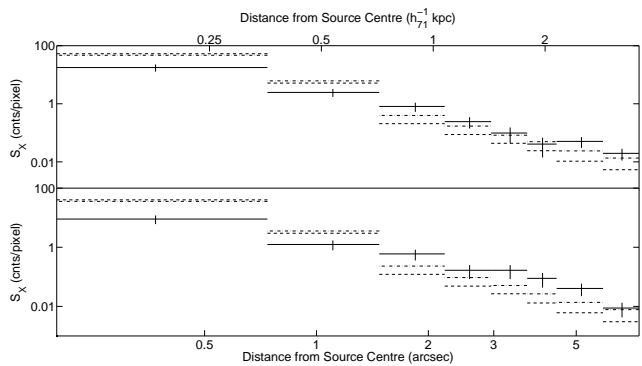}
\caption[NGC 1128 All Sources]
{Surface brightness profiles of sources in 3C 75/NGC 1128 compared to PSF. The dashed line is the 1 keV {\it ACIS} PSF and the dash-dot line is the 4 keV {\it ACIS} PSF.  The PSFs are normalised to the counts in the source.  The upper plot is {\it A400-42}'s profile, and the lower plot is {\it A400-43}'s profile.  We made similar profiles (not shown) for the northern and southern half of each source. The sources appear to be slightly extended overall and in both directions. \label{fig-3C75_SBProfile}}
\end{figure}

Figure-\ref{fig-A400-Zoom} shows a zoom in on the central region of Abell 400.  The nucleus in 3C 75 is clearly resolved into two sources.  The radio contour overlay shows that the radio and X-ray peak are co-spatial.  A comparison to the {\it ACIS} point spread function (PSF), shows both sources to be slightly extended (see Fig.-\ref{fig-3C75_SBProfile}).  To check the PSF model, we compared several other point sources in the field of view and found that they were consistent with the PSF.  To further test whether the extension was due to an interaction between the two sources we created four semi-annular profiles: the northern and southern half of {\it A400-42}, and the southern and northern half of {\it A400-43}.  These four profiles (not shown) indicate that the sources are indeed extended in both directions, and thus give no evidence of interaction between the sources.  

\begin{figure}
\includegraphics[width=85.0mm,angle=0]{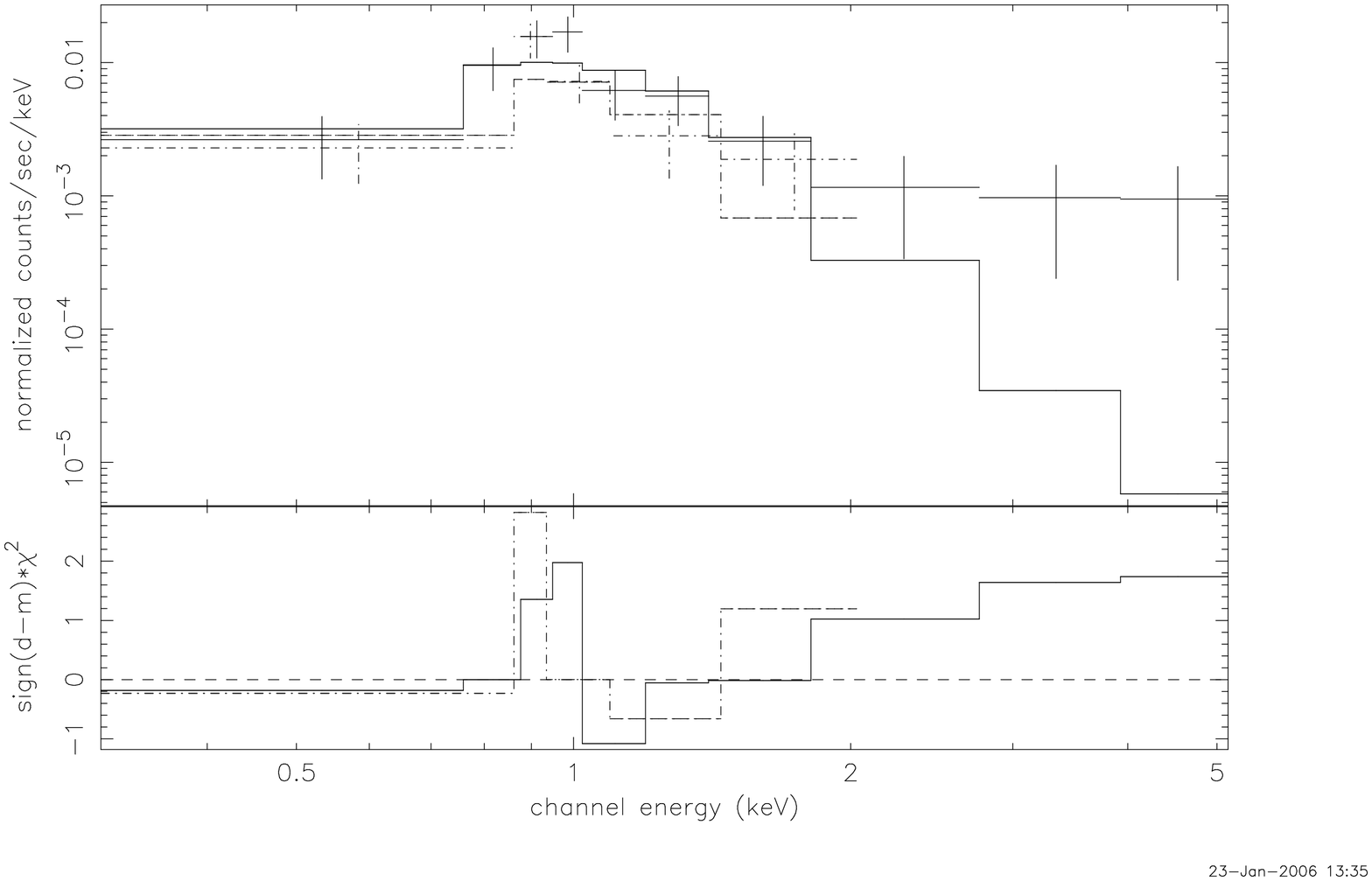}
\caption[Fits to {\it A400-42} and {\it A400-43}]
{Simple absorbed powerlaw fit to {\it A400-42} (solid) and {\it A400-43} (dash-dot).  In both cases there are significant residuals around 1 keV.  In addition {\it A400-42} has residuals above $\sim$2 keV.  Both spectra have been grouped to have at least 25 counts/bin.  For {\it A400-43}, there were fewer than 25 counts above $\sim$2.5 keV so that data was excluded from the fit. \label{PSfits}}
\end{figure}

In our {\it Chandra} observation {\it A400-42} and {\it A400-43} have 274 counts and 151 counts, respectively.  The spectrum of {\it A400-42} is much harder than that of {\it A400-43}.   Dividing the 0.3-10 keV spectra at 1.25 keV yields a hardness ratio of 0.97 for {\it A400-42} and 0.47 for {\it A400-43}.  This difference may be due to absorption at low energies by a dust column observed in {\it A400-42} and not found in {\it A400-43} \citep{sparks00}.  We extracted and fit spectra from each source.  For both sources a simple absorbed powerlaw fit to their spectra can be ruled out.  In the case of {\it A400-42}, reduced $\chi^2$ for the fit is $\sim$1.3, but significant residuals are visible at 1 kev and above $\sim$2 keV (see Fig.-\ref{PSfits}).  For {\it A400-42} there are also significant residuals around 1 keV and  reduced $\chi^2$ is $>2$.  For a simple absorbed thermal fit to the spectra, compared to the simple absorbed powerlaw, the residuals around 1 keV are fit for {\it A400-42}, but there are still significant residuals above $\sim$2 keV.  For {\it A400-43}, the spectrum is sufficiently fit to an absorbed thermal spectrum, with a temperature of 0.7$^{+0.3}_{-0.1}$ keV and metalicity fixed at solar values.  The residuals seen above 2 keV in the thermal model fit to the spectrum of {\it A400-43} are well modelled by the addition of a powerlaw component.  This model gives a cool thermal component $kT$ = 0.8$^{+0.3}_{-0.2}$ keV, a nonthermal component with a photon index of $\Gamma_{X}$ = 0.1$^{+2.9}_{-3.0}$, and an photoelectric absorption of $N_{H}< 20 \times 10^{20}$ cm$^{-2}$.   Considering the dust lane observed in {\it A400-42}, it seems strange that the 90\% upper limit is not much larger than the best fit value of the inner-most annulus ($N_H=14.14 \times 10^{20}$ cm$^{-2}$).

\subsection{Analysis of the Radio and X-ray Images}
The surface brightness map (see Fig.-\ref{fig-A400-SBMap}) clearly shows the elongation of the cluster gas to the southwest of the emission peak.  The emission peak (excluding the central AGN) is centred on the region containing the two AGN and is $\sim$1$\arcmin$.68 (or $\sim$49 $h_{71}^{-1}$ kpc) to the northeast of the EWC.  In Fig.-\ref{fig-A400-C1} a sudden drop in surface brightness is visible to the east of the emission peak, and seems to be coincident with the flaring and decollimation of the jets (see Sect.-\ref{ejet} for further discussion).  Additionally, this figure shows that on small scales, the jets can be seen bending to the northeast, aligned with the axis of elongation.  On larger scales (see Fig.-\ref{fig-A400-SBMap}), the extended radio emission seems to turn almost 90$^\circ$ in opposite directions.  

Figure-\ref{fig-A400-kTMap} shows the hardness implied temperature of the central region of the cluster.  The emission peak, centred on the two central AGN, appears to be a region of cool gas.  There appears to be a hotspot to the south, centred close to the EWC of the cluster.  The surrounding gas seems to be of an intermediate temperature.  Outside these regions, other features are also visible (such as a cool spot to the north), but these are not statistically significant.
As discussed below, we extracted and fit spectra for the three labelled regions in Fig.-\ref{fig-A400-kTMap}.

\begin{table*}
\caption{Spectral fit of total cluster emission on the I-Chips.}
\label{specfits}
\centering
\begin{tabular}{l c c c c c c c}
\hline
\hline
Model$^{\diamond}$ & $kT$   & $$Z                    & $kT_{2}$    & $N_{H}$     & Norm.($kT$)  & Norm.($kT_{2}$) & $\chi^{2}$/dof \\
       & (keV)& (Solar$^{\dagger}$)  &  (keV)      &  ($10^{20}$ cm$^{-2}$) & (10$^{-2}$ cm$^{-5}$) & (10$^{-2}$ cm$^{-5}$) &\\
\hline
APEC              & 2.35$^{+0.07}_{-0.07}$   & 0.57$^{+0.07}_{-0.07}$ &                          & 8.51$^{\ddagger}$       & 1.12$^{+0.04}_{-0.04}$ &                          & 480.2/478 \\
APEC              & 2.28$^{+0.10}_{-0.10}$   & 0.52$^{+0.08}_{-0.07}$ &                          & 9.97$^{+1.33}_{-1.31}$ & 1.16$^{+0.06}_{-0.06}$ &                          & 476.8/477 \\
APEC+APEC         & 2.68$^{+>61.32}_{-0.30}$ & 0.57$^{+0.08}_{-0.07}$ &  1.69$^{+0.49}_{-0.63}$  & 8.51$^{\ddagger}$       & 0.84$^{+0.26}_{-0.79}$ &  0.26$^{+0.84}_{-0.24}$& 474.3/476 \\
APEC+APEC         & 2.05$^{+0.09}_{-0.11}$ & 0.44$^{+0.08}_{-0.09}$ &  $>$2.46 & 10.91$^{+1.51}_{-1.44}$ & 1.15$^{+0.07}_{-0.07}$ &  0.10$^{+0.05}_{-0.05}$& 467.7/475 \\
\hline
\end{tabular}

\raggedright
$^{\diamond}$All models include photoelectric absorption with the Wisconsin cross-sections \citep{morrison}.\\
$^{\dagger}$We took solar abundances from \citet{anders89}.\\
$^{\ddagger}$Photoelectric absorption frozen at the galactic value.\\
\end{table*}

\subsection{The Overall Cluster Temperature}
Table-\ref{specfits} shows our fits to the overall cluster spectrum.  For our fit to the entire cluster spectrum, within 90\% confidence, the radio measured value of $N_H$ is rejected.  Additionally, our measured value of $N_H$ is consistent with the value obtained by \citet{ikebe02} of 9.38 $\times$ 10$^{20}$ cm$^{-2}$.  Furthermore,  \citet{baumgartner05} recently demonstrated that in regions with galactic column densities $\gtrsim 5 \times 10^{20}$ cm$^{2}$, the X-ray column densities are higher than found by radio measurements.  This made us suspicious that the radio value may be lower than the true Galactic value.  As discussed below, when analysing sections of the cluster, we found regions with high photoelectric absorption.  Therefore we adopt the single thermal model with free $N_H$ as our overall cluster model.  Adding a second thermal component only gives a marginal improvement to the fit (ftest$\sim$89.6\%) and  the high temperature component is unrestrained.  For a second thermal component and a free photoelectric absorption, the fit improves, but high temperature component has an unphysical best fit value of $>$64 keV.  Therefore, we report the overall cluster parameters as, a temperature (on the I-Chips) of $kT$ =2.28$^{+0.10}_{-0.10}$ keV, a metalicity of $Z$ = 0.52$^{+0.08}_{-0.07}$ solar, and a photoelectric absorption of $N_{H}$ = 9.97$^{+1.33}_{-1.31}$ $\times 10^{20}$ cm$^{-2}$.  The temperature value is between the two latest reported ${\it ASCA}$ values, $kT$ = 2.12$^{+0.06}_{-0.06}$ \citep{fukazawa04} and $kT$ = 2.43$^{+0.13}_{-0.12}$ \citep{ikebe02}.  We note that \citet{ikebe02} fit the data to a two temperature model and \citet{fukazawa04} fit the data excluding the core.  With {\it Chandra}, the central AGN can easily be excluded, and does not affect our temperature measurement.

\subsection{High, Spatially Varying Galactic Absorption}
\label{phoabs}

\begin{figure}
\includegraphics[width=85.0mm]{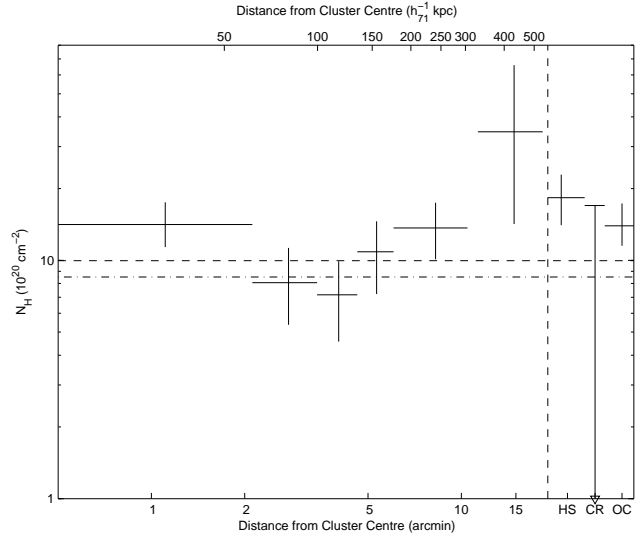}
\caption[A400: {\it ACIS} $N_H$ Profile]
{Photoelectric absorption ($N_H$) profile of Abell 400.  The dashed vertical black line seperates the profile values from the values of the three special regions.  The three special regions are labeled: HS - Hotspot, CR - Central Region, and OC - Outer Central Region.  See Fig.-\ref{fig-A400-kTMap} to see their positions.  The solid black lines represent the best fit projected values.  The dashed-dot line is the weight averaged radio value of 8.51 $\times$ 10$^{20}$ cm$^{-2}$ from the {\it RAIUB} Survey.  The dashed black line is the {\it Chandra} best fit value.  For the four regions consistent with the radio value, all subsequent tables and plots are for models with the column density fixed at the radio measured value.   \label{fig-A400-nHPlot}}
\end{figure}

As discussed above, we extracted six annuli centred at the {\it Chandra} EWC.  Since the overall cluster photoelectric absorption was higher than the average radio value, we fit the projected annuli with a free photoelectric absorption.  The results are shown in Fig.-\ref{fig-A400-nHPlot}.  The absorption seems to be high at the cluster centre and at the outer edge.  We only considered projected annuli, because fitting deprojected annuli with free $N_{H}$ can lead to misleading results.  This is due to the fact that if there is not a constant column density, the outer regions, when projected onto the inner regions, either over or under estimate the emission contributed to the central regions.  If the Galactic column density increases toward the centre, the outer regions projected onto the central regions will account for too many low energy photons emitted from the central region.  This leads to an even higher best fit value for $N_H$ and possible incorrect temperature measurement.  Likewise, if there is an outward increasing $N_H$, too few low energy photons are accounted for by the outer regions leading to a lower best fit temperature in the central region.  Once we determined the column density for the projected annuli, for all subsquent fits (unless otherwise noted) we used the radio measured value for three of the regions consistent with it (Annuli 2-4).  Since the CR (see Sect.-{\it Deprojection} above) is contained within the inner-most annulus and has a column density consistent with the inner-most region, we used that value rather than the radio value when fitting its spectrum.  Unless otherwise noted, we left the column density free for the regions that were not consistent with the radio value.  To deproject the annuli without over/underestimating the column density for the central regions, we co-fit the data allowing each annulus to have different foreground absorption.  This model is correct if the absorption is Galactic in origin, but spatially varying.

\begin{figure}
\includegraphics[width=85.0mm]{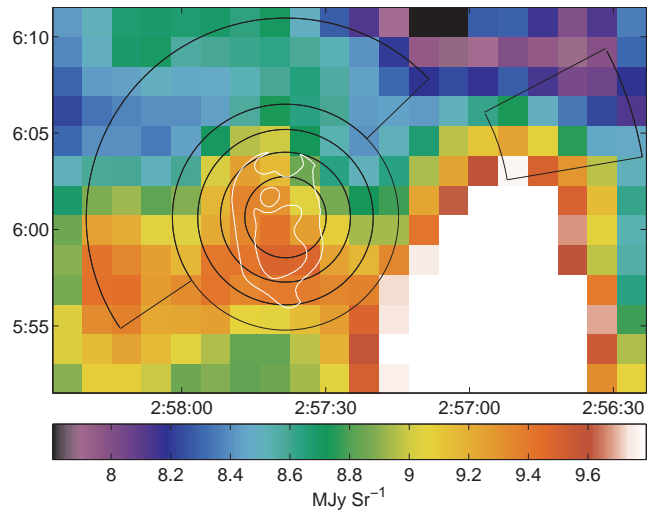}
\caption[A400: 100$\mu$ Emission]
{{\it IRAS} 100$\mu$ emission in the vicinity of A400 with the 6 annuli plus three special regions.  Note the peak in the emission coincident with the Hotspot region. (see Fig.-\ref{fig-A400-kTMap} for figure labels).  The colourscale is cut at 9.8 MJy Sr$^{-1}$ to show the variations within our Regions for A400.  The cut-off emission region to the southwest of our regions peaks at $\sim$12 MJy Sr$^{-2}$, which implies a column density of $\sim$14.4 $\times$ 10$^{20}$ cm$^{-2}$.  Although it is not coincident with our regions, it is within 20$\arcmin$ of the cluster centre and shows there may be column density gradients throughout this region.  \label{fig-A400-100uMap}}
\end{figure}

Additional evidence that Galactic column density in front of A400 is higher than the radio value is seen in the 100$\mu$ maps.  \citet{boulanger88} found a correlation between 100$\mu$ emission and hydrogen column density.  Plotting our regions over the 100$\mu$ map, we see that there is peak to the south of the EWC (see Fig.-\ref{fig-A400-100uMap}).  We further investigated this model by finding the average 100$\mu$ emission from each region and then estimating the column density from the relationship determined by \citet{boulanger88}. Our results are shown in Table-\ref{100uemission} .  We find that 100$\mu$ implied absorption column is higher than the radio implied value for all regions, but we find no strong gradient across the regions.  In two cases, the X-ray fit 90\% confidence value is lower than the 100$\mu$ value, whereas at the centre and outermost regions it is higher.  Since the scatter for the relationship between 100$\mu$ emission and $N_H$ is very high, the disagreement is not an important problem.  We present 100$\mu$ emission measurements simply to give more evidence that the column density may be higher in this region than given by the radio value.  Moreover, there appears to be a region of high 100$\mu$ emission in the vicinity of the cluster (see Fig.-\ref{fig-A400-100uMap}).  The 100$\mu$ emission in this region implies an $N_H$ column density consistent with the values ($N_H \sim$14 $\times$ 10$^{20}$ cm$^{-2}$) found in our inner and outer regions.  This gives further evidence that these type of $N_H$ gradients are possible in the region around A400.  Finally, we reiterate the results of \citet{baumgartner05}, who found that for $N_H > 5 \times 10^{20}$ cm$^{-2}$, the column density for X-ray measurements (so called $N_{HX}$) were higher than given by radio measurements ($N_H$).  Their explanation of this phenomenon is that high column densities, molecular as well as atomic hydrogen contributes to X-ray absorption, but is not included in radio measurements.   
\begin{table}
\caption{100$\mu$ emission for the annuli and regions in A400. Column 2 gives the implied $N_H$ using the relation between $N_H$ and 100$\mu$ emission found by \citet{boulanger88}.  All the 100$\mu$ implied values are above the radio value of 8.51 $\times$ 10$^{20}$ cm$^{-2}$.}  
\label{100uemission}
\centering
\begin{tabular}{lccc}
\hline
\hline

 Region  & $\langle I_{100\mu}\rangle$    &   Implied $N_{H}$  & Measured $N_{H}$  \\ 
         &    MJy/Sr         &       (10$^{20}$ cm$^{-2}$)  &        (10$^{20}$ cm$^{-2}$)   \\
\hline
01  & 9.256 & 11.11 & 11.37 - 17.51 \\
02  & 9.228 & 11.07 &  5.37 - 11.28 \\
03  & 9.100 & 10.92 &  4.56 -  9.91 \\
04  & 8.896 & 10.68 &  7.22 - 14.58 \\
05  & 8.669 & 10.40 & 10.10 - 17.46 \\
06  & 8.653 & 10.38 & 14.21 - 65.96 \\
01N & 9.173 & 11.01 &  9.61 - 18.73 \\
01S & 9.339 & 11.21 & 11.64 - 20.81 \\
02N & 9.097 & 10.92 &  6.76 - 15.46 \\
02S & 9.359 & 11.23 &  0.79 -  8.82 \\
HS  & 9.460 & 11.35 & 14.05 - 22.89 \\
CR  & 9.301 & 11.16 &  $<$16.99 \\
OC  & 9.240 & 11.09 & 11.52 - 17.33  \\

\hline
\end{tabular}

\raggedright
\end{table}

\subsection{Annular Analysis}
\begin{figure*}
\includegraphics[width=178.0mm]{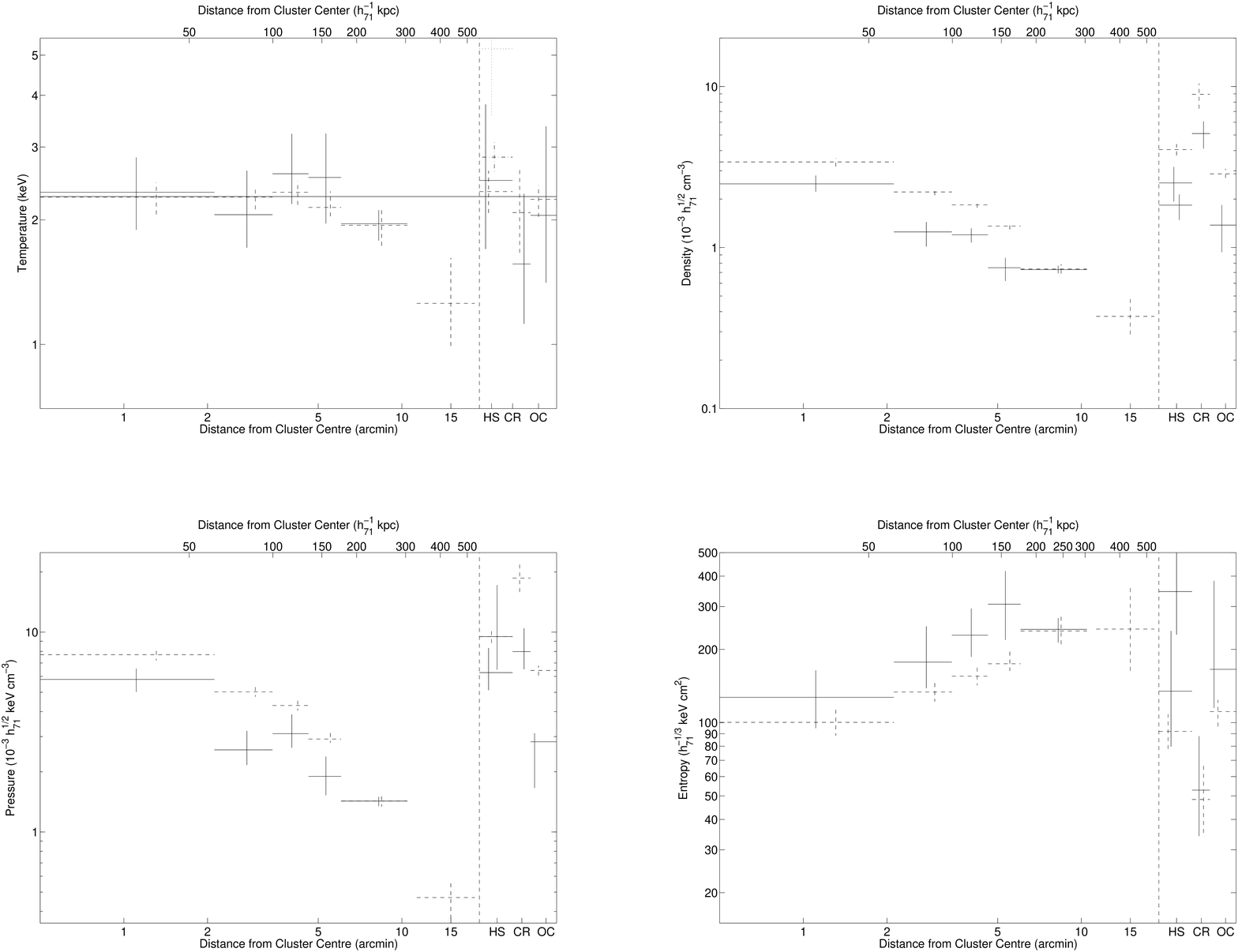}
\caption[A400: {\it ACIS} ALL Profile]
{Temperature Profile of Abell 400.  Solid black lines are deprojected temperatures.  The long, horizontal, solid line gives the average cluster temperature.  The dashed lines are the projected temperatures.  The vertical black line separates the annuli from the three special regions: HS-Hotspot, CR-Central Region, and OC-Outer Central Region.  The dash-dot line for HS is the \emph{projected} temperature with the photoelectric absorption fixed at the radio measured value.  The dotted line for HS gives the \emph{deprojected} temperature for the Hotspot with the photoelectric absorption fixed at the radio measured value.  Statistically it appears that the Hotspot on the temperature map is simply a increased column density.  The few counts in the CR made the deprojected fit very uncertain, therefore the {emp deprojected} fit plotted here is for the slightly larger BCR.  Based on the projected and deprojected results, {\it A400-42} and {\it A400-43} are in the lowest entropy region.\label{fig-A400-ALLPlot}}
\end{figure*}

As discussed above, we extracted spectra from six concentric annuli and three ``special'' regions.  The results of the projected and deprojected fits can be found in Table-\ref{project_annuli} and Table-\ref{deproject_annuli}, respectively.  A plot of selected projected and deprojected results can be seen in Fig.-\ref{fig-A400-ALLPlot}.  

Table-\ref{deproject_annuli} shows that although the best fit parameters of the CR make it stand-out from the other regions (e.g. lowest temperature, lowest entropy, highest pressure, etc...), the uncertainties are too large draw any conclusions.  By slightly extending the CR into the OC region, we were able to collect enough signal to constrain most of the parameters on the region.  We define this larger region as the Big Central Region (BCR).  In all parameters, except temperature, the BCR is not consistent with the OC region.  Since it contains part of the spectrum from the OC region, we argue that the best fit values for BCR are conservative estimates for the parameters in CR.    When plotting the \emph{deprojection} results in Fig.-\ref{fig-A400-ALLPlot} we used the parameters determined for BCR rather than CR.

\begin{table*}
\caption{Spectral fit of projected annuli and regions. We fit all models with both radio determined $N_{H}$ and free $N_{H}$.  We report values different from the radio determined value for those spectra which had a best fit value higher than the radio determined value, and when freeing $N_{H}$ gave a signficant improvement to the fit.}
\label{project_annuli}
\centering
\begin{tabular}{lccccccc}
\hline
\hline

 Radius/ & $kT$    &   $Z$              &  $N_{H}$   & Norm.     & $\rho$                   & Pressure                       & Entropy \\ 
 Region  & (keV) & (Solar$^{\dagger}$) & (10$^{20}$ cm$^{-2}$) & (10$^{-3}$ cm$^{-5}$) & (10$^{-3}$ $h_{71}^{1/2}$ cm$^{-3}$) & (10$^{-3}$ $h_{71}^{1/2}$ keV cm$^{-3}$)   &  ($h_{71}^{-1/3}$ keV cm$^{2}$)  \\
\hline
0 - 2$\arcmin$.12 & 2.33$^{+0.13}_{-0.27}$  & 0.46$^{+0.16}_{-0.14}$  & 14.14$^{+3.37}_{-2.77}$  & 2.29$^{+0.31}_{-0.24}$  & 3.40$^{+0.18}_{-0.21}$  & 7.93$^{+0.12}_{-0.71}$ & 103$^{+10}_{-15}$ \\
2$\arcmin$.12 - 3$\arcmin$.42 & 2.27$^{+0.15}_{-0.15}$  & 0.72$^{+0.21}_{-0.16}$  &  8.51$^{\ddagger}$        & 1.95$^{+0.17}_{-0.17}$  & 2.21$^{+0.09}_{-0.10}$  & 5.03$^{+0.28}_{-0.28}$ & 134$^{+12}_{-11}$ \\
3$\arcmin$.42 - 4$\arcmin$.61 & 2.33$^{+0.15}_{-0.16}$  & 0.62$^{+0.18}_{-0.14}$  &  8.51$^{\ddagger}$        & 2.07$^{+0.17}_{-0.17}$  & 1.84$^{+0.07}_{-0.07}$  & 4.29$^{+0.24}_{-0.25}$ & 155$^{+13}_{-13}$ \\
4$\arcmin$.61 - 6$\arcmin$.04 & 2.14$^{+0.20}_{-0.10}$  & 0.52$^{+0.19}_{-0.13}$  &  8.51$^{\ddagger}$        & 2.29$^{+0.19}_{-0.22}$  & 1.36$^{+0.05}_{-0.06}$  & 2.92$^{+0.21}_{-0.12}$ & 175$^{+21}_{-12}$ \\
6$\arcmin$.04 - 10$\arcmin$.46 & 1.94$^{+0.19}_{-0.21}$  & 0.26$^{+0.12}_{-0.09}$  &  13.67$^{+3.79}_{-3.57}$        & 3.23$^{+0.47}_{-0.40}$  & 0.73$^{+0.05}_{-0.04}$  & 1.43$^{+0.08}_{-0.09}$ & 238$^{+34}_{-34}$ \\
11$\arcmin$.30 - 18$\arcmin$.29 & 1.26$^{+0.36}_{-0.26}$  & 0.21$^{+0.58}_{-0.16}$  & 34.65$^{+31.31}_{-20.44}$  & 0.47$^{+0.35}_{-0.19}$  & 0.37$^{+0.12}_{-0.08}$  & 0.47$^{+0.08}_{-0.11}$ & 242$^{+115}_{-81}$ \\
HS  & 2.34$^{+0.29}_{-0.26}$  & 0.47$^{+0.26}_{-0.18}$  & 18.32$^{+4.57}_{-4.27}$  & 1.25$^{+0.21}_{-0.19}$  & 4.06$^{+0.33}_{-0.32}$  & 9.50$^{+0.69}_{-0.68}$ & 92$^{+16}_{-14}$ \\
HS   & 2.83$^{+0.26}_{-0.21}$  & 0.78$^{+0.32}_{-0.24}$  &  8.51$^{\ddagger}$        & 0.95$^{+0.10}_{-0.10}$  & 3.59$^{+0.18}_{-0.19}$  & 10.19$^{+0.86}_{-0.73}$ & 121$^{+14}_{-12}$ \\
CR & 2.08$^{+0.56}_{-0.42}$  & 0.36$^{+0.80}_{-0.29}$  &  14.14$^{\diamond}$        & 0.17$^{+0.06}_{-0.06}$  & 8.94$^{+1.50}_{-1.66}$  & 18.64$^{+3.75}_{-2.77}$ & 48$^{+20}_{-13}$ \\
CR & 2.35$^{+0.61}_{-0.47}$  & 0.49$^{+0.98}_{-0.39}$  &  8.51$^{\ddagger}$        & 0.14$^{+0.05}_{-0.05}$  & 8.30$^{+1.33}_{-1.51}$  & 19.51$^{+4.03}_{-3.26}$ & 57$^{+22}_{-15}$ \\
BCR  & 1.95$^{+0.25}_{-0.25}$  & 0.21$^{+0.24}_{-0.14}$  &   14.14$^{\diamond}$        & 0.49$^{+0.09}_{-0.09}$  & 6.79$^{+0.60}_{-0.62}$  & 13.26$^{+1.28}_{-1.26}$ & 54$^{+10}_{-9}$ \\
OC  & 2.24$^{+0.19}_{-0.21}$  & 0.63$^{+0.21}_{-0.20}$  & 13.96$^{+3.37}_{-2.44}$  & 1.97$^{+0.31}_{-0.22}$  & 2.87$^{+0.22}_{-0.16}$  & 6.43$^{+0.37}_{-0.37}$ & 111$^{+13}_{-15}$ \\

\hline
\end{tabular}

\raggedright
$^{\dagger}$We took solar abundances from \citet{anders89}.\\
$^{\ddagger}$Photoelectric absorption frozen at the radio value.\\
$^{\diamond}$Frozen at best fit value of inner-most annulus.\\
\end{table*}

\begin{table*}
\caption{Deprojection fit to annuli and regions.  We did not fix $N_H$ at the radio value for the same spectra as discussed in Table-\ref{project_annuli}.  The 11$\arcmin$.30-18$\arcmin$.29 region lies on the S2 Chip and was not included in the deprojection model.  The largest annulus on the I-Chips (6$\arcmin$.04 - 10$\arcmin$.46) is not deprojected in the sense that no emission was considered outside of it.  It was, however, co-fit with the other regions, so that the it was marginally affected by the fits to the inner regions.}
\label{deproject_annuli}
\centering
\begin{tabular}{lccccccc}
\hline
\hline

 Radius/ & $kT$    &   $Z$              &  $N_{H}$   & Norm.     & $\rho$                   & Pressure                       & Entropy \\ 
 Region  & (keV) & (Solar$^{\dagger}$) & (10$^{20}$ cm$^{-2}$)  & (10$^{-3}$ cm$^{-5}$) & (10$^{-3}$ $h_{71}^{1/2}$ cm$^{-3}$) & (10$^{-3}$ $h_{71}^{1/2}$ keV cm$^{-3}$)   &  ($h_{71}^{-1/3}$ keV cm$^{2}$)  \\
\hline
0 - 2$\arcmin$.12 & 2.33$^{+0.50}_{-0.44}$  & 0.25$^{+0.33}_{-0.19}$  & 14.37$^{+2.80}_{-2.84}$  & 1.25$^{+0.34}_{-0.25}$  & 2.49$^{+0.32}_{-0.27}$  & 5.80$^{+0.77}_{-0.78}$ & 127$^{+37}_{-32}$ \\
2$\arcmin$.12 - 3$\arcmin$.42 & 2.06$^{+0.57}_{-0.35}$  & 0.88$^{+1.32}_{-0.47}$  &  8.51$^{\ddagger}$        & 0.96$^{+0.31}_{-0.33}$  & 1.25$^{+0.18}_{-0.23}$  & 2.57$^{+0.63}_{-0.41}$ & 177$^{+71}_{-39}$ \\
3$\arcmin$.42 - 4$\arcmin$.61 & 2.58$^{+0.64}_{-0.40}$  & 0.70$^{+0.56}_{-0.36}$  &  8.51$^{\ddagger}$        & 1.71$^{+0.34}_{-0.34}$  & 1.20$^{+0.11}_{-0.12}$  & 3.10$^{+0.77}_{-0.46}$ & 229$^{+66}_{-42}$ \\
4$\arcmin$.61 - 6$\arcmin$.04 & 2.53$^{+0.70}_{-0.57}$  & 1.29$^{+1.44}_{-0.64}$  &  8.51$^{\ddagger}$        & 1.34$^{+0.44}_{-0.42}$  & 0.74$^{+0.11}_{-0.12}$  & 1.90$^{+0.49}_{-0.37}$ & 307$^{+113}_{-88}$ \\
6$\arcmin$.04 - 10$\arcmin$.46 & 1.96$^{+0.15}_{-0.18}$  & 0.28$^{+0.10}_{-0.08}$  & 13.48$^{+3.60}_{-3.19}$  & 10.48$^{+1.21}_{-1.04}$  & 0.73$^{+0.04}_{-0.04}$  & 1.43$^{+0.07}_{-0.08}$ & 242$^{+27}_{-28}$ \\
HS & 2.49$^{+1.31}_{-0.79}$  & 0.31$^{+1.16}_{-0.31}$  & 18.71$^{+4.99}_{-4.43}$  & 0.49$^{+0.29}_{-0.20}$  & 2.52$^{+0.65}_{-0.59}$  & 6.27$^{+2.05}_{-1.15}$ & 135$^{+104}_{-55}$ \\
HS  & 5.18$^{+4.08}_{-1.59}$  & 0.91$^{+1.28}_{-0.91}$  &  8.51$^{\ddagger}$        & 0.24$^{+0.09}_{-0.08}$  & 1.83$^{+0.30}_{-0.35}$  & 9.50$^{+7.66}_{-3.01}$ & 346$^{+278}_{-116}$ \\
CR  & 1.83$^{+1.57}_{-0.93}$  & 0.17$^{+3.20}_{-0.17}$  &  13.84$^{+3.39}_{-2.74}$$^{\diamond}$        & 0.08$^{+0.10}_{-0.06}$  & 6.22$^{+2.96}_{-2.92}$  & 11.39$^{+7.01}_{-3.45}$ & 54$^{+82}_{-33}$ \\
BCR  & 1.56$^{+0.58}_{-0.44}$  & 0.03$^{+0.22}_{-0.03}$  & 14.40$^{+3.34}_{-2.65}$$^{\diamond}$  & 0.28$^{+0.12}_{-0.10}$  & 5.10$^{+0.97}_{-0.99}$  & 7.99$^{+2.46}_{-1.46}$ & 53$^{+35}_{-19}$ \\ 
OC  & 2.05$^{+1.32}_{-0.64}$  & 0.46$^{+2.02}_{-0.41}$  & 13.07$^{+3.64}_{-3.01}$  & 0.46$^{+0.36}_{-0.25}$  & 1.38$^{+0.46}_{-0.44}$  & 2.83$^{+0.29}_{-1.17}$ & 166$^{+217}_{-51}$ \\

\hline
\end{tabular}

\raggedright
$^{\dagger}$We took solar abundances from \citet{anders89}.\\
$^{\ddagger}$Photoelectric absorption frozen at the radio value.\\
$^{\diamond}$Tied to the absorption in the inner-most region.\\
\end{table*}

The most striking feature of the temperature profile is that within 90\% confidence, all the annuli (both projected and deprojected), except the outer-most region, are consistent with each other.  Likewise, all but the outer two annuli are consistent with the best fit overall cluster temperature ($kT = 2.28$ keV).  Even the ``special regions'' are consistent with the overall cluster temperature and the five inner annuli.  The one exception is the deprojected $HS$ with $N_H$ frozen at the radio value (see Sect.-\ref{disc} for a discussion on the HS).  The lack of a central cooling region is consistent with the merger scenario, since mergers are thought to destroy cooling cores \citep{buote96}.  We also note an interesting detail, that on large scales the cluster temperature profile drops.  This behaviour has seen in general for clusters by e.g. \citet{degrandi02}, and most recently by \citet{vikhlinin05}.  Annulus 6, is the only annulus with a temperature not consistent with the other annuli. 

Whereas, the temperature profile and surface brightness map resemble a merging cluster, the projected density, pressure, and entropy profiles are more reminiscent of a relaxed cluster.  The central region is the highest density, highest pressure, lowest entropy region.  The pressure and density steadily drop in the outer regions, and the entropy rises until finally levelling off in the outer regions.  However, after examining the deprojected annuli, we find that the profiles do not resemble a relaxed cluster.  Specifically, there seems to be a sudden drop in density and pressure from the first to the second annulus, followed by a levelling between annulus 2 and 3.  There is also a levelling off of the density and pressure in between annulus 4 and annulus 5.  This, however, is due to projection effects.  That is, for the purposes of deprojection, we consider annulus 5 to be the outer region.  Since there is clearly emission beyond annulus 5, the density in this annulus is over-estimated.  A better comparison is the projected components when comparing annulus 4 and annulus 5.

Examining the special regions we find more evidence of merging in the cluster.  The CR is located $\sim$1$\arcmin$.7 from the EWC of the cluster and contains the densest, and lowest entropy gas.  Along with the HS, the CR has the highest pressure gas.  The HS is roughly located at the cluster's EWC, but has a temperature consistent with the rest of the cluster, if not higher.  It has a high pressure, but this seems to be due to a high temperature, rather than a high density.  This picture, as we discuss below, is consistent with a shock region.  The best fit lowest temperature is found in the CR, although with the current observation, it is not well constrained.  

\section{Discussion}
\label{disc}
Both nuclei of 3C 75 are clearly resolved and detected by {\it Chandra}.  Both sources are extended and cannot be adequately fit with a simple absorbed powerlaw model.  This picture is consistent with a central AGN providing the non-thermal emission and the surrounding galaxy providing the extended thermal emission.  For both {\it A400-42} and {\it A400-43} we find a temperature component consistent with galactic emission ($kT \sim 1$ keV) and much lower than any gas found in the cluster.  Many authors \citep[e.g.][]{beers92} have conjectured that {\it A400-42} and {\it A400-43} are the remnants of two dominant galaxies that have formed a dumbbell galaxy when their respective clusters merge.  The extended emission seen in X-rays is consistent with this picture, since it suggests a massive galaxy.  The suggested existence of this extended cool thermal emission in both cores indicates that thermal conduction from the much hotter intracluster gas may be suppressed, similar to the two central Coma Cluster galaxies \citep{vikhlinin01}.

As seen with {\it Einstein} and reported by \citet{beers92}, the cluster is elongated along the northeast-southwest direction.  This axis lines up with the initial bending of the jets and implies that on scales of up to tens of kpc, their bending is due to the motion of 3C 75 relative to the ICM.  Although \citet{owen85} argued from the kinematic data of \citet{hintzen82} that this could not be the case, the later analysis of the galaxy kinematics by \citet{beers92} demonstrated that this is a plausible explanation.

The sound speed in the cluster (using the average cluster temperature) is $\sim$850 km s$^{-1}$.  Based on the simulations by \citet{yokosawa85} the relative motion of the gas and the jets would need to be $\sim$1100 km s$^{-1}$ to cause the jets to bend.  This slightly supersonic motion should cause weak shocks in the gas.  Using the adiabatic jump shock conditions \citep[e.g.][]{landau59}, a shock with Mach number of ${\cal M} \sim$1.3 would compress the gas by a factor of $\sim$1.4.  This would heat the gas by a factor of 1.29.  Using the global cluster temperature ($kT = 2.28$ keV), this shock would heat the gas to a temperature of $\sim$2.9 keV.  This temperature is not ruled out by our measurements.  The merger velocity predicted by \citet{beers92} is $\sim$2000 km s$^{-1}$, or a Mach number of ${\cal M} \sim 2.35$.  This would produce a shock that would heat the gas to $kT = 5.8$ keV.  This temperature is only achieved for the fit to the deprojected HS with $N_{H}$ frozen at the radio value. 

Based on the position of the HS region and its parameters when $N_H$ is frozen at the radio value, it is an ideal candidate for a shock.  It has a high temperature ($kT_{\rm best fit} = 5.2$ keV) and entropy ($S_{\rm best fit} = 346 h_{71}^{-1/3}$ keV cm$^{2}$).  On the other hand, with $>$99\% confidence the high absorption model is preferred over the high temperature model.  As we discussed in Sect.-\ref{phoabs}, there also appears to be a peak in the 100$\mu$ emission (albeit a small one) over this region.  Is it really possible that nature has conspired against us to put a very high column density in the very position we expect a shock?  Moreover, the temperature increase is exactly what is expected for a weak shock of ${\cal M}$ $\sim$ 2.  The deprojected fit to the HS region produces a very high column density.  It is in fact, much higher than the first two annuli which contain it.  It also has a higher value than implied by any 100$\mu$ emission within the region.  One possibility we consider, is that the HS region suffers from a high column density and a high temperature.  In this case, the column density cannot be constrained and the temperature is pushed to an artifically low value.  In fact, within 90\% confidence, a high temperature is not ruled out ( $kT_{\rm max} \sim 3.80$ ), nor is an absorption parameter consistent with inner-most annulus ($N_{H_{{\rm min}}}$$\sim$14.28 $\times$ 10$^{20}$ cm$^{-2}$).  If we fit the deprojected model for the HS region and freeze its absorption parameter at the best fit value to the inner annulus ($N_{H}=$14.37 $\times$ 10$^{20}$ cm$^{-2}$), we obtain a best fit temperature of $\sim$3.5 keV.  The increase of $\chi^{2}$ is moderate, $\sim$2 for 3 degrees of freedom, giving an ftest likelihood of $\sim$88\% that the model with free $N_H$ is preferred.   This  lowers even further if we tie the absorption parameters between the HS and inner-most annulus (ftest = 78\%).  In this case the best fit temperature is $kT\sim3.1$ keV.  Using the bestfit overall cluster temperature, both these models give a shock temperature consistent with a weak shock of Mach number ${\cal M}\sim$1.4-1.5.  Given the sound speed of the cluster, this would give a relative velocity of $\sim$1200 km s$^{-1}$.  Given the position of the shock and the two AGN, it makes sense they are in the dense, low entropy region moving through the cluster causing the shock.  Taking the deprojected pressure of CR, it is $P$ $>$ 7.94 $h_{71}^{1/2}$ keV cm$^{-3}$, whereas the average pressure in the inner-most region is $P = 5.0-6.6$ $h_{71}^{1/2}$ keV cm$^{-3}$.  If this core is moving through the central region this suggests a Mach number ${\cal M} > 1.1$, with a best fit value of ${\cal M}$ = 1.7.  This argument is further strengthened when the CR is compared to the OC, which surrounds it.  The pressure of the gas in the OC is only $P$ = 1.6-3.1 $h_{71}^{1/2}$ keV cm$^{-3}$.  When this pressure is compared to the pressure in the CR, it gives an estimated Mach number of ${\cal M}$$\sim$2-3, consistent with the merger velocity estimated by \citet{beers92}.

We note that the above temperature analysis is for the best fit values rather than the range allowed by the errors.  For the former model with $N_H$ frozen at 14.37 $\times$ 10$^{20}$ cm$^{-2}$, the one-sigma errors allow a temperature as low as 2.8 keV, corresponding to a Mach number of ${\cal M}$$\sim$1.2.   For the latter model, the 1-$\sigma$ temperature can be as low as 2.5 keV, for a Mach number of ${\cal M}$$\sim$1.  Only in the case in which the photoelectric absorption is frozen at the radio value is a temperature consistent with a 2000 km s$^{-1}$ merger allowed.  Of course the core does not have to have a local relative velocity equal to the merger velocity.  It is quite possible that there are shocks consistent with 2000 km s$^{-1}$ and that they are simply not bright enough for us to observe.  We argue that the most likely scenario is that we are looking at gas shock heated as the core moves through at $\sim1200$ km s$^{-1}$.  Unfortunately, this shock region happens to be in a region (consistent with the inner annulus) with a high Galactic column density, making the temperature difficult to constrain.  We point out that the pressure difference between the CR and OC as further evidence of shocked gas.  With a longer observation, we should be able to disentangle the hot temperature and high Galactic column density and give a more precise measurement of the temperature in this region.

\subsection{Interaction of the Jets and the ICM}
In Fig.-\ref{fig-A400-Zoom} we present a high resolution map of the jets in 3C 75 at 4.5 GHz.  As seen in \citet{owen85}, both AGN are clearly visible and separated.  The jets brighten before widening into the radio lobes seen on the large scale.  The reason for the transition from jet to lobe is still not well understood, but it is thought that it may relate to a sudden change in the ICM pressure \citep[see e.g.]{Wiita92}.  To check for sudden density changes in the ICM at the point of flaring, we made surface brightness profiles from pie sections along each projected jet.  To the north both jets either entwine or overlap in projection.  To the east the two jets are separate until they both bend north and entwine, or overlap.

\subsubsection{The case of {\it A400-42}'s eastern jet}
\label{ejet} 
\begin{figure*}
\includegraphics[width=178.0mm,angle=0]{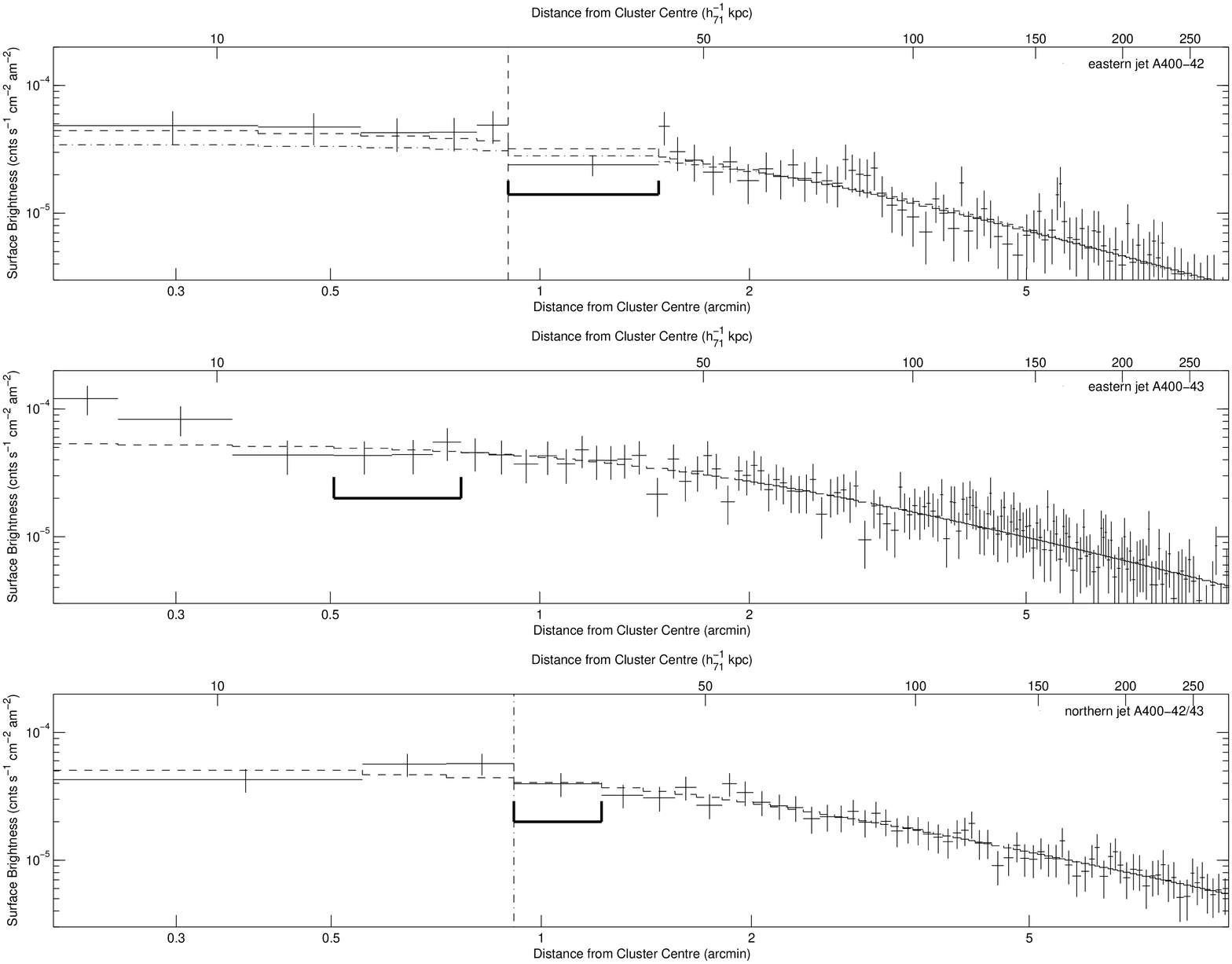}
\caption[Eastern (Northern) Jet]
{Surface Brightness profiles along the axis of the jets.  The thick "$U$'s" indicate the approximate point of flaring/transition from collimated jets to radio lobes.  The top is the eastern jet of {\it A400-42}. the middle is the eastern jet of {\it A400-43}, and the bottom is the combined jets to the north (see labels).  The dashed vertical line in the top figure shows the position of sudden drop in surface brightness coincident with the flaring of the eastern jet of {\it A400-42}.  Unfortunately this is also the position of a chip gap, and the drop might be an instrumental effect.  The dashed horizontal line gives the best fit $\beta$ model to all the points, and the dash-dot line in the top plot gives the best fit $\beta$ model for the points at radii equal to larger than the sudden drop.  Note that the sudden drop is not consistent with a $\beta$-model fit to all the surface brightness points, but is consistent with the latter.  The other two profiles are consistent with the former model as the point of flaring.  The vertical dashed-dot in the bottom plot shows the approximate position of the chip gap in that profile.  There is a drop in the surface brightness, but it is not as dramatic or statistically significant as for {\it A400-42}'s eastern jet (top plot). \label{figJSBP}}
\end{figure*}

{\it A400-42}'s eastern jet flares and expands at a point where the {\it ICM}'s density seems to drop quickly (see Fig.-\ref{figJSBP}).  Unfortunately this region coincides with a chip gap, so that the sudden drop may simply be due to an incorrect exposure correction.  Note that the combined northern jet also crosses the chip gap at a normal to it, and although the surface brightness drops slightly at this point (see Fig.-\ref{figJSBP}), it is not as statistically significant as is the case for the {\it A400-42}'s eastern jet.  

To further check the significance of the drop in surface brightness in {\it A400-42}'s eastern jet, we fit a $\beta$-model to all the surface brightness profiles along the jets.  In all cases, except {\it A400-42}'s eastern jet, the $\beta$-model was consistent with the surface brightness at the point of flaring.  To quantify the drop in density at the point {\it A400-42}'s eastern jet flares, we fit only the points in the surface brightness after and including the point of the drop.  This model, along with the fit which includes the inner regions, can be seen in Fig.-\ref{figJSBP}.  Note that when all points are used, the point of the drop in surface brightness is not consistent with the model, but when the inner points are removed, it is.  It is difficult to get a precise measurement of the density in this region because of the uncertainty in modelling the projection effects.  However, as a rough calculation, we used our $\beta$-model fit to the outer points to account for the projected emission.  We find a drop in density of a factor of $\sim$4.2.  To check the significance of the drop, we calculated the expected drop between the two regions for our best fit $\beta$-model.  The expected drop in density is a factor of $\sim$1.1.   This means that our density drop is $\sim$3.8$\times$ greater than expected for typical density profile. 

It is unclear why only {\it A400-42}'s eastern jet shows this behaviour.  The fact that the others don't may mean it is only a coincidence, or perhaps the projected geometry is such that it is easier to detect the drop than in the other jets.  Most likely, is that there is more than one condition that can lead to the decollimation of the jets.  The fact, however, that the drop coincides with a chip gap makes us suspicious that it may simply be an instrumental effect.  A follow-up {\it Chandra} or {\it XMM-Newton} observation where this point does not fall in a gap should be sufficient to clarify this point.

Finally, we point out, that the sudden rise in surface brightness after the drop, is coincident with the galaxy {\it A400-41} (also PGC 011193). {\it A400-41} is an elliptical galaxy about 1$\arcmin$.5 from 3C 75 that shows no radio emission.  The actual galaxy itself is identified as an X-ray source by {\it wavdetect}, but even after the point source removal, excess emission is still visible in the smoothed image.  It is interesting, is that the radio emission seems to bend suddenly and go around {\it A400-41} (see Fig.-\ref{fig-A400-C1}).  

Since {\it A400-41} is not detected in radio, and is rather bright in X-rays, we assume that the emission is thermal (i.e. it is not an active galaxy).  We used our $\beta$-model fit for the cluster emission in the direction of {\it A400-41} to subtract off the projected emission.  Assuming the gas in this galaxy had a temperature of 0.5 keV and solar metalicity,   we obtain an average density of $n_{\rm ion} = 1.4 \times 10^{-2}$ $h_{71}^{1/2}$ cm$^{-3}$.  For a temperature of 0.5 keV, this gives an average pressure of $P_{\rm gal} = 1.2 h_{71}^{1/2} \times 10^{-11}$ dynes cm$^{-2}$.  

Simulations by \cite{Wang00} showed that it was possible for galaxies to deflect jets under certain conditions. To see if it is possible for {\it A400-41} to deflect the jet, we can roughly estimate the minimum speed a jet can have so the pressure it exerts on a galaxy is less than the galaxy's average pressure.  That is,
\begin{equation}
P_{\rm gal} > \frac{\rho v^2}{2} = \frac{\dot{E}}{vA},
\label{Peqn}
\end{equation}
 where $P_{\rm gal}$ is the pressure of the galaxy's gas, $\rho$ and $v$ are the jet's density and velocity respectively, $A$ is the cross section area of the jet, and  $\dot{E}$ = $L$/$\epsilon$, where $\epsilon$$\sim$0.01 is the assumed radiative efficiency.  We can very roughly determine the luminosity of the jet from the radio observations.  For {\it A400-41}'s eastern jet, we obtain a flux of $\sim$ 2.5 Jy and $\sim$  0.44 Jy at 330 MHz and 4.5 GHz respectively.  This suggests a spectral index of $\alpha \sim$ -0.7.  Integrating from 0 to 4.5 GHz, we obtain an integrated flux of 4 $\times 10^{-13}$ ergs cm$^{-2}$ s$^{-1}$.  Taking the luminosity distance of the cluster $D_{L}$ = 105.0 Mpc, this gives a luminosity of $L \sim  5 \times 10^{41} h_{71}^{-2}$ ergs s$^{-1}$, or a kinetic luminosity of $\dot{E}$ $\sim 5 \times 10^{43} h_{71}^{-2}$ ergs s$^{-1}$.   Solving for $v$ gives:
 \begin{equation}
 v > \frac{\dot{E}}{A P_{\rm gal}}.
 \label{veqn}
 \end{equation}
From the 4.5 GHz radio map, we estimate the width of the jet to be $\sim$ 10$\arcsec$, which gives a cross-sectional area of $A \sim 4 \times 10^{44} h_{71}^{-2}$ cm$^{2}$.  Plugging into equation-\ref{veqn} gives $v$ $\gtrsim 10^{5}$ km s$^{-1}$.  Therefore, it is possible for {\it A400-41} to deflect the jet if it is a light, supersonic jet. Incidently, examining Fig-\ref{fig-A400-C1} it appears as if {\it A400-41} is not only deflecting {\it A400-42}'s eastern jet, but moving through the jets pushing {\it A400-43}'s eastern jet into {\it A400-42}'s eastern jet.  However, with no apparent X-ray tail, or shock heating to the north east of {\it A400-41} it is impossible to make any quantitative arguments for this scenario.

\subsubsection{ICM Gradient}
\citet{beers92} were the first authors to point out that A400 may be a merging cluster.  Additionally they claimed that the {\it Einstein} image shows elongation along the same axis as the jet bending.  Based on these two facts, they concluded that the bending of the jets could be related to relative motion of the ICM.  With {\it Chandra}'s imaging capability, it is now possible to see clearly that A400 is elongated along the same direction as the bending of the jets (e.g. see Fig.-\ref{fig-A400-C1}).  Does that necessarily mean that the bending of the extended radio emission is related to bulk flow of the ICM?  

\begin{figure}
\includegraphics[width=85.0mm,angle=270]{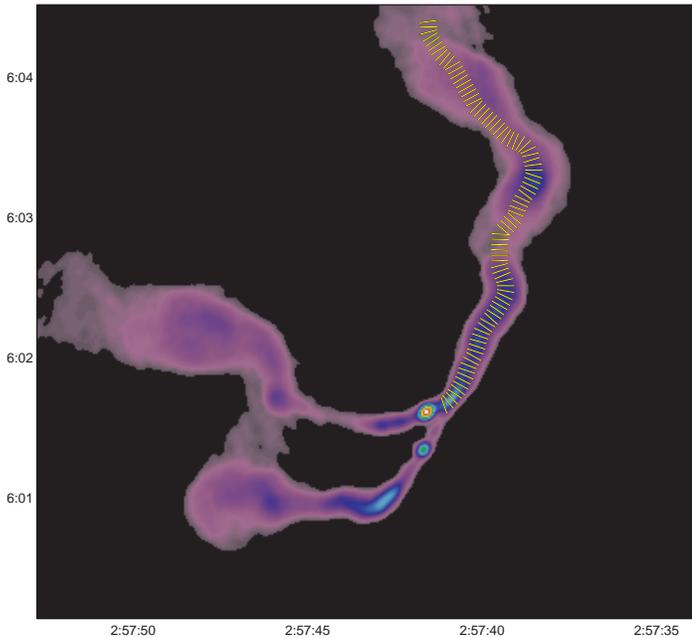}
\caption[Trace of northwestern jet]
{This is a 4.5 GHz image of 3C 75.  The bars across the northern jet represent our algorithm's prediction of the jet centre at that point and the normal to the direction of motion (in the plane of the sky).  It is clear that the algorithm is not perfect at all points, but does manage to follow the path of the jet quite reasonably for the majority of points. \label{figNJpath}}
\end{figure}

It is difficult to model the direction of the motion of the jet and the pressure gradient of the ICM.  To first order, we can model the motion of the jet across the plane of the sky to see if it follows the ICM gradient, as expected for a buoyant jet.  To do this, we started at the each AGN and moved 10 pixels (7$\arcsec$.5) to the left (or right).  We took a cross-section of our jet perpendicular to the x-axis (masking out the other jet) and calculated the emission weighted centre of the cross-section.  That is, the y-coordinate of the emission weighted centre along the cross-section.  Considering the AGN to be the origin, the  angle of the line connecting the AGN to the emission-weighted centre of the cross-section is arctan($y_{\rm ewc}$/10), where $y_{\rm ewc}$ is the y-coordinate of the emission-weighted centre.  However, because of fluctuations in the jet's width and luminosity, the arbitrarily chosen initial angle can affect the rotation angle calculated.  Therefore, we rotated our coordinate system by the calculated angle (arctan($y_{\rm ewc}$/10)) and repeated the procedure using the X-axis of the rotated coordinate system.   We repeated this calculation until the change in calculated rotation angle was less than 5$^{\circ}$.  We considered the net angle (from the original X-axis) to be the direction of the jet from the origin and the point 10-pixels down the jet.  We then restarted the procedure considering our new point to be the origin, our final X-axis (from our previous iteration) to be starting X-axis on the current iteration, and moving 3 pixels (2$\arcsec$.25) along the X-axis.  We repeated the procedure described above, but with 3 pixel intervals rather than 10 pixels.  The reason for the difference in the number of pixels is that the initial large step of 10 pixels was simply needed to clear the AGN before starting the iteration.  Figure-\ref{figNJpath} shows the normals to the direction at the calculated points for the northern jet.  It is clear the algorithm is not perfect, but averaged over a few points it gives a good indication of the projected direction of the jet.

\begin{figure}
\includegraphics[width=85.0mm,angle=270]{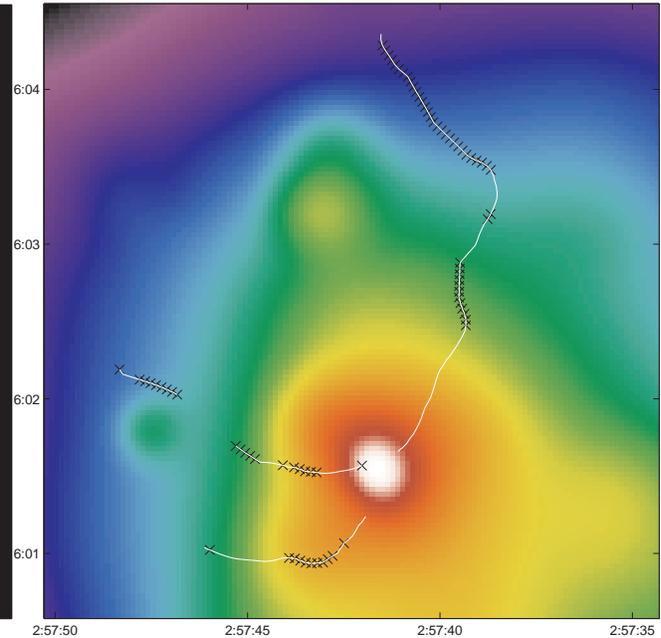}
\caption[3C 75 Iteration with ICM]
{This image shows our smoothed A400 surface brightness map.  The white lines represent our algorithm's estimation of the position of the jet (see text).  For the two eastern jets, the algorithm was unable to follow the point where the southern jet expands, turns north and appears to interact with the northern jet.  Even without our algorithm it is clear in this region that the jet does not follow the surface brightness gradient (see Fig.-\ref{fig-A400-C1}).  The black "x's" represent points where the jet and the surface brightness gradient differed by more than 20$^{\circ}$. \label{fig3C75iteraction}}
\end{figure}

\begin{figure}
\includegraphics[width=85.0mm,angle=0]{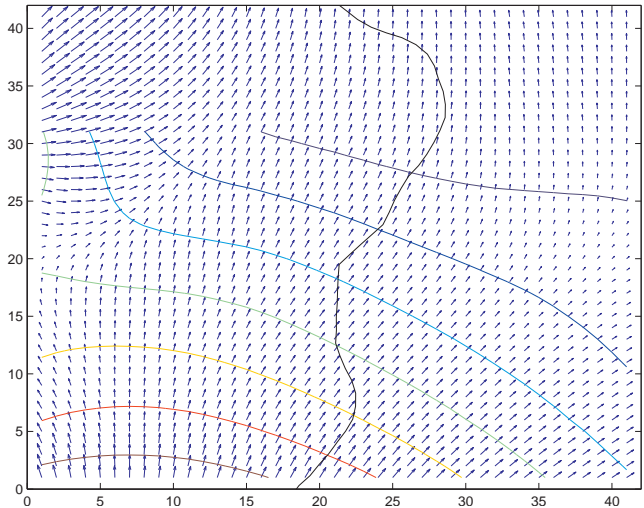}
\caption[Northwest Jet direction versus ICM gradient]
{This image shows the surface brightness inverse gradient versus a section of the northern jet.  The blue arrows represent the direction of the surface brightness inverse gradient, with the size of the arrow being proportional to its magnitude at that point.  The black line shows the path of the jet.  The jet seems to follow the gradient and then abruptly turn across it.  It if were following the gradient, we would expect a shift in the gradient to coincide or precede a bend in the jet. \label{figICMvsjetgradarrows}}
\end{figure}

We numerically calculated the inverse gradient of the smoothed surface brightness image of the ICM, and compared it to the projected path of the jets.  As mentioned above, the algorithm for determining the jet direction was not perfect, so if only a few points differed, we would attribute that to inaccurate modelling by the algorithm.  Figure-\ref{fig3C75iteraction} shows the smoothed surface brightness image of A400, with the algorithms estimation of the jet's position.  The "x's" indicate any position that the jet and ICM inverse gradient differed by more than 20$^{\circ}$.  Perhaps what is most surprising in this image is, not only do the jets not follow the ICM gradient, but that the locations where the jets bend correspond to places where the direction of the ICM gradient and the jets diverge.  Although there are many places the jets and ICM gradient are in the same direction, this is expected.  The jets are going out of the cluster system and on the large scale the surface brightness decreases with distance from the cluster centre.  However, in places where the jets bend, one would expect the ICM gradient to bend if the jet were following it.  Instead we find that jets bend across the ICM gradient.  Figure-\ref{figICMvsjetgradarrows} shows this for the northern jet.  The blue arrows in this diagram show the direction of the inverse surface brightness gradient, with their length being proportional to the strength of the gradient.  The black line shows the path of the northern jet.  The jet is initially following the gradient and then suddenly turns across it, before turning back in the parallel direction, but ultimately cuts back across the gradient.

\begin{figure}
\includegraphics[width=85.0mm,angle=0]{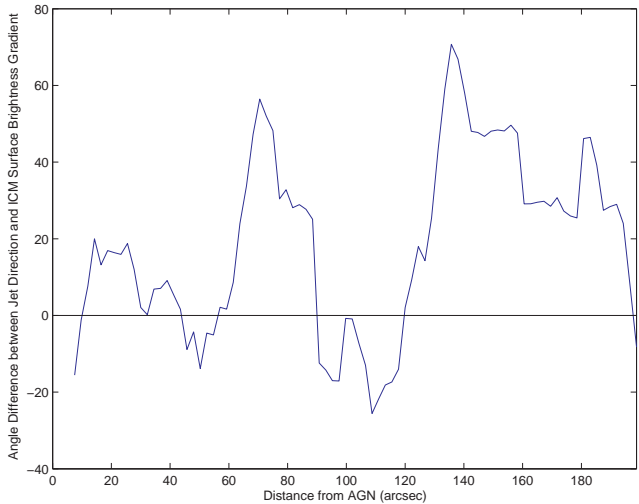}
\caption[Northwest Jet direction versus ICM gradient]
{This figure plots the angle between the surface brightness gradient and the jet direction for the northern jet.  It is representative of the other two jet as well.  If the jet were following the gradient, but wandering around it, we would expect a sine like structure centred on y=0.  Rather we find the jet spends most of its time turning in the positive direction. \label{figICMvsjetgrad}}
\end{figure}

We considered the possibility that perhaps the jet was meandering along the gradient, like a marble going from side to side as it rolls down a gutter.  We plotted the angle between the jet and the surface brightness inverse gradient.  If the angle swung between positive and negative in a sinusoidal fashion, we could argue that the jet is following the ICM gradient, but its inertia carries it around the direction of the local path.  Fig.-\ref{figICMvsjetgrad} shows the results for the northern jet, which is typical of all three.  There appears to be no correlation and the jet seems to mostly turn in one direction relative to the ICM gradient.

We argue that this demonstrates that the jet does not bend because it follows the gradient in the ICM as one would expect for a buoyant jet.  The ICM wind, which would be parallel to the plane of the sky relative to the projected jets is impossible to measure directly, but seems the most likely candidate for causing the initial bending of the jets.  The sudden turns of the jets after decollimation may be due to turbulence of the ICM which is less directed than the wind caused by the AGN moving through the ICM.

Turning to the large scale interactions (see Fig.-\ref{fig-A400-SBMap}),  we see that the jets turn almost 90$^\circ$, away from each other and normal to the elongation axis.  A consistent picture with the above arguments, is that this extended radio emission has not been affected by the wind.  That is, if we consider the jets initially pointed to the east and west before the merger, then as the core moves through the ICM, the jets become bent to the north.  The result is that the outer regions of the extended radio emission still point east and west, but the are connected to the inner regions that point north.  Based on this simple model, we can estimate the merger time, using the projected angle of 18$^\circ$ estimated by \citet{beers92} and our core velocity of $\sim$ 1200 km s$^{-1}$.  We estimate the distance to the bends in the jets to be $\sim$130 $h_{71}^{-1}$ kpc away from the central region.  At that distance, travelling at 1200 km s$^{-1}$, it would take the AGN $\sim$10$^{8}$ years to reach their current position.  This time period is consistent with the estimated dynamical age of the tails ($\sim10^{8}$ years) \citep{odea85}.  Given this short time scale, it suggests that the cores have just merged, consistent with \citet{beers92} prediction that the clusters were merging at a low angle (with respect to the plane of the sky) and were caught as they were crossing; rather than a line of sight merger where the clusters overlap in projection during the merger process.  Comparison with simulations done by \citet{roettiger97} suggests that the mass ratio of the clusters is 1:4 or 1:8 and that the cores are between $\sim$0-0.5 Gyr from passing.  Since both jets initially bend to the northeast, this suggests that the proto-{\it SMBBH} is a remnant of a previous merger and was not produced in the current merger.  This suggests that the two AGN belong to the cluster coming in from the northeast.  In this scenario, the jets, initially straight and in opposite directions, are being bent by the ICM from the current merger with a cluster to the southwest.  

\section{Conclusion}
The proto-{\it SMBBH}, 3C 75, is clearly separated in our {\it Chandra} images.  Both {\it A400-42} and {\it A400-43} show evidence of extended emission and cannot be fit with a simple absorbed powerlaw model.  For {\it A400-43}, we can fit the spectrum to an absorbed, cool ($kT < 1$), thermal component.  In the case of {\it A400-42} we have enough signal to fit a combined thermal plus non-thermal model.  For this model we find a $<$1keV thermal component and an unconstrained ($\Gamma$ = -2.9 - 3.0) non-thermal component.

We find strong evidence that A400 is a merging cluster.  The {\it Chandra} image shows clear elongation along the northeast-southwest axis.  There is no visible temperature gradient in the central regions. 

We find possible interaction between {\it A400-42}'s eastern jet and ICM.  It seems to flare and decollimate at a point where the ICM surface brightness drops by a factor of $\sim$4.2.  This behaviour has been observed in simulations by \citet{Wiita92} with similar density drops.  We acknowledge that it is possible this drop may be an instrumental effect, as it occurs in a chip gap.  With another {\it Chandra} or with an {\it XMM} observation, we should easily be able to confirm this drop.  {\it A400-42}'s eastern jet, also seems to take a sharp northern turn and travel around {\it A400-41}.  The average density of {\it ISM} in {\it A400-41} is large enough to bend that particular jet if it is a light, relativistic jet.  

In terms of the other jets, we see no such behaviour.  This could be due to projection effects making the detections in {\it A400-42}'s eastern jet easier.  More likely is that there is more than one process that accounts for bending, flaring and decollimation.  For instance, \citet{norman88} found that a strong galactic wind could decollimate a jet.  Also, {\it A400-41} may be responsible for one bend in {\it A400-42}'s eastern jet, but there is no evidence of clouds or galaxies causing the other bends.

To further study the interaction of the jets and the gas, we modelled the direction of the jet and compared it to the negative gradient of the surface brightness.  We found that in general the jet did not follow the gradient, and even more so, we found that sudden bends in the jet took it across gradients.  This suggests that the path of the jets is not determined by buoyancy and the ICM wind and turbulence is the most likely culprit for the bending of the jets on scales of 10's of kpc.

We have some evidence for a shocked region to the southwest of the cluster core.  If the photoelectric absorption is frozen at the best fit value for the 
inner-most annulus, the best fit deprojected temperature in the HS region is consistent with a shock of Mach number ${\cal M}$$\la$1.4.  At the speed of sound in this cluster, this gives a velocity of $\sim$1200 km s$^{-1}$.  In this case the most likely cause for the initial bending of the jets would be the local ICM wind.  Moreover the shock region is found to the south of the dense, low entropy core containing the two AGN.  If that core is moving through the cluster and is responsible for the shock, then the local wind would be $\sim$1200 km s$^{-1}$.  The fact that the jets all bend in a similar direction, suggest that they must have the same relative velocity to the ICM (i.e. travelling in the same direction relative to the wind).  The proto-{\it SMBBH} must, therefore, be a remnant of a previous merger.  This makes an unbound (i.e. projected) system unlikely.  

\begin{acknowledgements}
The authors wish to thank to Wendy Lane for providing the 330 MHz radio map of 3C 75, and Bill Forman for helpful early discussions.

C. L. S., T.H.R, and T.E.C were supported in part by by the National Aeronautics and Space Administration through $Chandra$ Award GO4-5132X, issued by the $Chandra$ X-ray Observatory Center, which is operated by the Smithsonian Astrophysical Observatory for and on behalf of NASA under contract NAS8-39073, and by NASA {\it XMM-Newton} Grant NNG05GO50G.

T.H.R. and D.S.H. achnowledge support from the Deutsche Forschungsgemeinschaft through Emmy Noether research grant RE 1462.

\end{acknowledgements}

\label{lastpage}

\end{document}